\documentclass[a4paper, 11pt]{article}

\usepackage{amsfonts}
\usepackage{amssymb}
\usepackage{bbm}
\usepackage{epstopdf}

\usepackage{hyperref}

\usepackage{amsmath}
\usepackage{amsfonts}
\usepackage{amssymb}

\newcount\driver
\newcount\bozza

\let\a=\alpha \let\b=\beta  \let\g=\gamma  \let\d=\delta \let\e=\varepsilon
\let\z=\zeta  \let\h=\eta   \let\th=\theta \let\k=\kappa \let\l=\lambda
\let\m=\mu    \let\n=\nu    \let\x=\xi     \let\p=\pi    \let\r=\rho
\let\s=\sigma \let\t=\tau   \let\f=\varphi \let\ph=\varphi\let\c=\chi
\let\ps=\psi  \let\y=\upsilon \let\o=\omega\let\si=\varsigma
\let\G=\Gamma \let\D=\Delta  \let\Th=\Theta\let\L=\Lambda \let\X=\Xi
\let\P=\Pi    \let\Si=\Sigma \let\F=\Phi    \let\Ps=\Psi
\let\O=\Omega \let\Y=\Upsilon

\def\\{\hfill\break} \let\==\equiv
\let\txt=\textstyle\let\dis=\displaystyle

\let\io=\infty 

\let\0=\noindent\def\pagina{{\vfill\eject}}
\def\bra#1{{\langle#1|}}\def\ket#1{{|#1\rangle}}
\def\media#1{{\langle#1\rangle}}
\def\ie{{i.e. }}\def\eg{{e.g. }}
\let\dpr=\partial \def\der{{\rm d}} \let\circa=\cong

\def\tende#1{\,\vtop{\ialign{##\crcr\rightarrowfill\crcr
 \noalign{\kern-1pt\nointerlineskip} \hskip3.pt${\scriptstyle
 #1}$\hskip3.pt\crcr}}\,}
\def\circage{\lower2pt\hbox{$\,\buildrel > \over {\scriptstyle \sim}\,$}}
\def\otto{\,{\kern-1.truept\leftarrow\kern-5.truept\to\kern-1.truept}\,}

\def\PPP{{\cal P}}\def\EE{{\cal E}}\def\MM{{\cal M}}
\def\CC{{\cal C}}\def\FF{{\cal F}} \def\HHH{{\cal H}}\def\WW{{\cal W}}
\def\TT{{\cal T}}\def\NN{{\cal N}} \def\BBB{{\cal B}}\def\III{{\cal I}}
\def\RR{{\cal R}}\def\LL{{\cal L}} \def\JJ{{\cal J}} \def\OO{{\cal O}}
\def\DD{{\cal D}}\def\AAA{{\cal A}}\def\GG{{\cal G}} \def\SS{{\cal S}}
\def\KK{{\cal K}}\def\UU{{\cal U}} \def\QQ{{\cal Q}} \def\XXX{{\cal X}}

\def\T#1{{#1_{\kern-3pt\lower7pt\hbox{$\widetilde{}$}}\kern3pt}}
\def\VVV#1{{\VV #1}_{\kern-3pt
\lower7pt\hbox{$\widetilde{}$}}\kern3pt\,}
\def\W#1{#1_{\kern-3pt\lower7.5pt\hbox{$\widetilde{}$}}\kern2pt\,}

\def\lis{\overline}
 
  \def\sign{{\rm sign}\,}
\def\indica{\leaders \hbox to 0.5cm{\hss.\hss}\hfill}
\def\guida{\leaders\hbox to 1em{\hss.\hss}\hfill}

\def\hh{{\bf h}} \def\HH{{\bf H}} \def\AA{{\bf A}} \def\qq{{\bf q}}
\def\VV{{\cal V}}
\def\BB{{\bf B}}  \def\PP{{\bf P}} \def\pp{{\bf p}}
\def\vv{{\bf v}} \def\xx{{\bf x}} \def\yy{{\bf y}} \def\zz{{\bf z}}
\def\hhh{{\bf h}}
\def\ii{{\bf i}}\def\kk{{\bf k}}

\def\VV#1{{\,\underline#1\,}}
\def\ul{\underline}

\def\defin{{\buildrel def\over=}}
\def\wt{\widetilde}
\def\wh{\widehat}

\mathchardef\dd   = "050E
\mathchardef\aa   = "050B
\mathchardef\bb   = "050C
\mathchardef\ggg  = "050D
\mathchardef\xxx  = "0518
\mathchardef\zzzzz= "0510
\mathchardef\oo   = "0521
\mathchardef\lll  = "0515
\mathchardef\mm   = "0516
\mathchardef\Dp   = "0540
\mathchardef\H    = "0548
\mathchardef\FFF  = "0546
\mathchardef\ppp  = "0570
\mathchardef\Bn   = "0517
\mathchardef\pps  = "0520
\mathchardef\fff  = "0527
\mathchardef\FFF  = "0508
\mathchardef\nnnnn= "056E

\def\to{\rightarrow}
\def\la{\left\langle}
\def\ra{\right\rangle}

\def\qed{\hfill\raise1pt\hbox{\vrule height5pt width5pt depth0pt}}
\def\hf#1{{\hat \f_{#1}}}
 \def\tg#1{{\tilde g_{#1}}}

\def\Val{{\rm Val}}
\def\indic{\hbox{\raise-2pt \hbox{\indbf 1}}}

\def\RRR{\hbox{\msytw R}} \def\rrrr{\hbox{\msytww R}}
\def\rrr{\hbox{\msytwww R}} \def\CCC{\hbox{\msytw C}}
\def\cccc{\hbox{\msytww C}} \def\ccc{\hbox{\msytwww C}}
\def\NNN{\hbox{\msytw N}} \def\nnnn{\hbox{\msytww N}}
\def\nnn{\hbox{\msytwww N}} \def\ZZZ{\hbox{\msytw Z}}
\def\zzzz{\hbox{\msytww Z}} \def\zzz{\hbox{\msytwww Z}}

\def\TTT{\hbox{\msytw T}} \def\tttt{\hbox{\msytww T}}
\def\ttt{\hbox{\msytwww T}}

\def\ul#1{{\underline#1}}

\def\V0{{\bf 0}}
\def\defi{\,{\buildrel def\over=}\,}

\font\tenmib=cmmib10 
\font\sevenmib=cmmib7\font\fivemib=cmmib5

\font\ottorm=cmr8

\mathchardef\Ba   = "050B  
\mathchardef\Bb   = "050C  
\mathchardef\Bg   = "050D  
\mathchardef\Bd   = "050E  
\mathchardef\Be   = "0522  
\mathchardef\Bee  = "050F  
\mathchardef\Bz   = "0510  
\mathchardef\Bh   = "0511  
\mathchardef\Bthh = "0512  
\mathchardef\Bth  = "0523  
\mathchardef\Bi   = "0513  
\mathchardef\Bk   = "0514  
\mathchardef\Bl   = "0515  
\mathchardef\Bm   = "0516  
\mathchardef\Bn   = "0517  
\mathchardef\Bx   = "0518  
\mathchardef\Bom  = "0530  
\mathchardef\Bp   = "0519  
\mathchardef\Br   = "0525  
\mathchardef\Bro  = "051A  
\mathchardef\Bs   = "051B  
\mathchardef\Bsi  = "0526  
\mathchardef\Bt   = "051C  
\mathchardef\Bu   = "051D  
\mathchardef\Bf   = "0527  
\mathchardef\Bff  = "051E  
\mathchardef\Bch  = "051F  
\mathchardef\Bps  = "0520  
\mathchardef\Bo   = "0521  
\mathchardef\Bome = "0524  
\mathchardef\BG   = "0500  
\mathchardef\BD   = "0501  
\mathchardef\BTh  = "0502  
\mathchardef\BL   = "0503  
\mathchardef\BX   = "0504  
\mathchardef\BP   = "0505  
\mathchardef\BS   = "0506  
\mathchardef\BU   = "0507  
\mathchardef\BF   = "0508  
\mathchardef\BPs  = "0509  
\mathchardef\BO   = "050A  
\mathchardef\BDpr = "0540  
\mathchardef\Bstl = "053F  

\def\bT{{\bf T}}

\def\V#1{{\bf#1}}
\let\aa=\Ba\let\fff=\Bf\let\defin=\defi\def\HHH{{\cal H}}
\def\VV{{\cal V}}

\let\wt=\widetilde\def\AAA{{\cal A}}\let\oo=\Bo\let\nn=\Bn
\let\pps=\Bps\def\hhh={\V h}
\let\bb=\Bb

\def\RRR{\hbox{\msytw R}} \def\rrrr{\hbox{\msytww R}}
\def\rrr{\hbox{\msytwww R}} \def\CCC{\hbox{\msytw C}}
\def\cccc{\hbox{\msytww C}} \def\ccc{\hbox{\msytwww C}}
\def\NNN{\hbox{\msytw N}} \def\nnnn{\hbox{\msytww N}}
\def\nnn{\hbox{\msytwww N}} \def\ZZZ{\hbox{\msytw Z}}
\def\zzzz{\hbox{\msytww Z}} \def\zzz{\hbox{\msytwww Z}}
\def\TTT{\hbox{\msytw T}} \def\tttt{\hbox{\msytww T}}
\def\ttt{\hbox{\msytwww T}}

\let\ul=\underline\def\hh{{\V h}}

\def\Tr{{\rm Tr}}

\def\ins#1#2#3{\vbox to0pt{\kern-#2 \hbox{\kern#1 #3}\vss}\nointerlineskip}
\newdimen\xshift \newdimen\xwidth \newdimen\yshift
\newcount\griglia

\def\insertplot#1#2#3#4#5#6{%
\begin{figure}[h]
\begin{center}
\vspace{#2pt}
\begin{minipage}{#1pt}
#3
\ifnum\driver=1
\griglia=#6
\ifnum\griglia=1
\openout13=griglia.ps
\write13{gsave .2 setlinewidth}
\write13{0 10 #1 {dup 0 moveto #2 lineto } for}
\write13{0 10 #2 {dup 0 exch moveto #1 exch lineto } for}
\write13{stroke}
\write13{.5 setlinewidth}
\write13{0 50 #1 {dup 0 moveto #2 lineto } for}
\write13{0 50 #2 {dup 0 exch moveto #1 exch lineto } for}
\write13{stroke grestore}
\closeout13
\includegraphics{griglia.ps}\fi
\includegraphics{#4.ps}\fi
\ifnum\driver=2
\fi
\end{minipage}
\end{center}
\caption{#5}
\end{figure}
}

\def\gtopl{\hbox{\msxtw \char63}}
\def\ltopg{\hbox{\msxtw \char55}}

\newdimen\shift \shift=-1truecm
\def\lb#1{%
\ifnum\bozza=1
\label{#1}\rlap{\kern\shift{$\scriptstyle#1$}}
\else\label{#1}
\fi}

\def\be{\begin{equation}}
\def\ee{\end{equation}}
\def\bea{\begin{eqnarray}}\def\eea{\end{eqnarray}}
\def\bean{\begin{eqnarray*}}\def\eean{\end{eqnarray*}}
\def\bfr{\begin{flushright}}\def\efr{\end{flushright}}
\def\bc{\begin{center}}\def\ec{\end{center}}
\def\ba#1{\begin{array}{#1}} \def\ea{\end{array}}
\def\bd{\begin{description}}\def\ed{\end{description}}
\def\bv{\begin{verbatim}}\def\ev{\end{verbatim}}
\def\nn{\nonumber}
\def\Halmos{\hfill\vrule height10pt width4pt depth2pt \par\hbox to \hsize{}}
\def\pref#1{(\ref{#1})}
 
\def\virg{\quad,\quad}

\newtheorem{lemma}{Lemma}[section]
\newtheorem{theorem}{Theorem}[section]

\newdimen\xshift \newdimen\xwidth \newdimen\yshift \newdimen\ywidth

\def\ins#1#2#3{\vbox to0pt{\kern-#2\hbox{\kern#1 #3}\vss}\nointerlineskip}

\def\eqfig#1#2#3#4#5{
\par\xwidth=#1 \xshift=\hsize \advance\xshift
by-\xwidth \divide\xshift by 2
\yshift=#2 \divide\yshift by 2
\line{\hglue\xshift \vbox to #2{\vfil
#3 \includegraphics{#4.ps}
}\hfill\raise\yshift\hbox{#5}}}

\def\8{\write12}

\begin{document}

\let\a=\alpha \let\b=\beta  \let\g=\gamma  \let\d=\delta \let\e=\varepsilon
\let\z=\zeta  \let\h=\eta   \let\th=\theta \let\k=\kappa \let\l=\lambda
\let\m=\mu    \let\n=\nu    \let\x=\xi     \let\p=\pi    \let\r=\rho
\let\s=\sigma \let\t=\tau   \let\f=\varphi \let\ph=\varphi\let\c=\chi
\let\ps=\Psi  \let\y=\upsilon \let\o=\omega\let\si=\varsigma
\let\G=\Gamma \let\D=\Delta  \let\Th=\Theta\let\L=\Lambda \let\X=\Xi
\let\P=\Pi    \let\Si=\Sigma \let\F=\Phi    \let\Ps=\Psi
\let\O=\Omega \let\Y=\Upsilon

\def\PPP{{\cal P}}\def\EE{{\cal E}}\def\MM{{\cal M}} \def\VV{{\cal V}}
\def\CC{{\cal C}}\def\FF{{\cal F}} \def\HHH{{\cal H}}\def\WW{{\cal W}}
\def\TT{{\cal T}}\def\NN{{\cal N}} \def\BBB{{\cal B}}\def\III{{\cal I}}
\def\RR{{\cal R}}\def\LL{{\cal L}} \def\JJ{{\cal J}} \def\OO{{\cal O}}
\def\DD{{\cal D}}\def\AAA{{\cal A}}\def\GG{{\cal G}} \def\SS{{\cal S}}
\def\KK{{\cal K}}\def\UU{{\cal U}} \def\QQ{{\cal Q}} \def\XXX{{\cal X}}

\def\kk{{ k}} 
\def\qq{{ q}} \def\pp{{\bf p}}\def\Ve{{\bf e}}
\def\vv{{\bf v}} \def\xx{{x}} \def\yy{ {y}} \def\zz{{ z}}
\def\aa{{\bf a}}\def\hh{{\bf h}}\def\kk{{k}}
\def\mm{{\bf m}}\def\PP{{\bf P}}\def\dd{{\boldsymbol{\d}}}

\def\nn{\nonumber}

\def\RRR{\hbox{\msytw R}} \def\rrrr{\hbox{\msytww R}}
\def\rrr{\hbox{\msytwww R}} \def\CCC{\hbox{\msytw C}}
\def\cccc{\hbox{\msytww C}} \def\ccc{\hbox{\msytwww C}}
\def\NNN{\hbox{\msytw N}} \def\nnnn{\hbox{\msytww N}}
\def\nnn{\hbox{\msytwww N}} \def\ZZZ{\hbox{\msytw Z}}
\def\zzzz{\hbox{\msytww Z}} \def\zzz{\hbox{\msytwww Z}}


\def\\{\hfill\break}
\def\={:=}
\let\io=\infty
\let\0=\noindent\def\pagina{{\vfill\eject}}
\def\media#1{{\langle#1\rangle}}
\let\dpr=\partial
\def\sign{{\rm sign}}
\def\const{{\rm const}}
\def\tende#1{\,\vtop{\ialign{##\crcr\rightarrowfill\crcr\noalign{\kern-1pt
    \nointerlineskip} \hskip3.pt${\scriptstyle #1}$\hskip3.pt\crcr}}\,}
\def\otto{\,{\kern-1.truept\leftarrow\kern-5.truept\to\kern-1.truept}\,}
\def\defin{{\buildrel def\over=}}
\def\wt{\widetilde}
\def\wh{\widehat}
\def\to{\rightarrow}
\def\la{\left\langle}
\def\ra{\right\rangle}
\def\qed{\hfill\raise1pt\hbox{\vrule height5pt width5pt depth0pt}}
\def\Val{{\rm Val}}
\def\ul#1{{\underline#1}}
\def\lis{\overline}
\def\V#1{{\bf#1}}
\def\be{\begin{equation}}
\def\ee{\end{equation}}
\def\bea{\begin{eqnarray}}
\def\eea{\end{eqnarray}}
\def\nn{\nonumber}
\def\pref#1{(\ref{#1})}
\def\ie{{\it i.e.}}
\def\lb{\label}
\def\eg{{\it e.g.}}
\def\sl{{\displaystyle{\not}}}
\def\Tr{\mathrm{Tr}}
\def\eu{\mathrm{e}}

\newcommand{\sgn}{\text{sgn}}

\newcount\driver

\font\cs=cmcsc10 scaled\magstep1 \font\ottorm=cmr8 scaled\magstep1
\font\msxtw=msbm10 scaled\magstep1 \font\euftw=eufm10
scaled\magstep1 \font\msytw=msbm10 scaled\magstep1
\font\msytww=msbm8 scaled\magstep1 \font\msytwww=msbm7
scaled\magstep1 \font\indbf=cmbx10 scaled\magstep2
\font\bigteni=cmmi10 scaled \magstep2 \font\bigtenrm=cmr10 scaled
\magstep2 \font\grbold=cmmib10 scaled\magstep1 \font\amit=cmmi7

\def\sf{\textfont1=\amit}

{\count255=\time\divide\count255 by 60
\xdef\hourmin{\number\count255}
        \multiply\count255 by-60\advance\count255 by\time
   \xdef\hourmin{\hourmin:\ifnum\count255<10 0\fi\the\count255}}

\let\a=\alpha \let\b=\beta    \let\g=\gamma     \let\d=\delta     \let\e=\varepsilon
\let\z=\zeta  \let\h=\eta     \let\th=\vartheta \let\k=\kappa     \let\l=\lambda
\let\m=\mu    \let\n=\nu      \let\x=\xi        \let\p=\pi        \let\r=\rho
\let\s=\sigma \let\t=\tau     \let\f=\varphi    \let\ph=\varphi   \let\c=\chi
\let\ps=\psi  \let\y=\upsilon \let\o=\omega     \let\si=\varsigma
\let\G=\Gamma \let\D=\Delta   \let\Th=\Theta    \let\L=\Lambda    \let\X=\Xi
\let\P=\Pi    \let\Si=\Sigma  \let\F=\Phi       \let\Ps=\Psi
\let\O=\Omega \let\Y=\Upsilon

\def\PP{{\cal P}}\def\EE{{\cal E}}\def\MM{{\cal M}}\def\VV{{\cal V}}
\def\CC{{\cal C}}\def\FF{{\cal F}}\def\HH{{\cal H}}\def\WW{{\cal W}}
\def\TT{{\cal T}}\def\NN{{\cal N}}\def\BB{{\cal B}}\def\ZZ{{\cal Z}}
\def\RR{{\cal R}}\def\LL{{\cal L}}\def\JJ{{\cal J}}\def\QQ{{\cal Q}}
\def\DD{{\cal D}}\def\AA{{\cal A}}\def\GG{{\cal G}}\def\SS{{\cal S}}
\def\OO{{\cal O}}\def\XXX{{\bf X}}\def\YYY{{\bf Y}}\def\WWW{{\bf W}}
\def\KK{{\cal K}}\def\LLL{{\bf L}}

\def\AAA{{\cal A}}

\def\RRR{\mathbb{R}} \def\ZZZ{\mathbb{Z}}  
\def\rrrr{\hbox{\msytww R}}
\def\rrr{\hbox{\msytwww R}}       \def\CCC{\mathbb{C}} 
\def\cccc{\hbox{\msytww C}}       \def\ccc{\hbox{\msytwww C}}
\def\NNN{\hbox{\msytw N}}         \def\nnnn{\hbox{\msytww N}}
\def\nnn{\hbox{\msytwww N}}       \def\ZZZ{\hbox{\msytw Z}}
\def\zzzz{\hbox{\msytww Z}}       \def\zzz{\hbox{\msytwww Z}}
\def\TTT{\mathbb{T}}  
\def\tttt{\hbox{\msytww T}}
\def\ttt{\hbox{\msytwww T}}

\def\bfs{{\bf s}}\def\bfe{{\bf e}}
\def\pp{{\bf p}}\def\qq{{\bf q}}\def\ii{{\bf i}}\def\xx{{x}}
\def\aa{{\bf a}}\def\cc{{\bf c}}\def\bb{{\bf b}} \def\dd{{\bf d}}
\def\yy{{y}}\def\kk{{k}}\def\mm{{\bf m}}\def\nn{{\bf n}}
\def\zz{{\bf z}}\def\uu{{\bf u}}\def\vv{{\bf v}}\def\ww{{\bf w}}
\def\tt{{\bf t}}\def\gg{{\bf g}}\def\hh{{\bf h}}\def\rr{{\bf r}}
\def\bT{{\bf T}}\def\bP{{\bf P}}\def\ff{{\bf f}}\def\bA{{\bf A}}
\def\xxi{\hbox{\grbold \char24}}
\def\pp{{p}}

\def\oo{{\underline \omega}} \def\usp{{\underline s}} \def\ub{{\underline b}}
\def\us{{\underline \sigma}}       \def\uo{{\underline \omega}}
\def\un{{\underline \nu}}          \def\ul{{\underline \lambda}}
\def\um{{\underline \mu}}          \def\ux{{\underline\xx}}
\def\uk{{\underline \kk}}          \def\uq{{\underline\qq}}
\def\ual{{\underline \a}}         \def\ubb{{\underline\bb}}
\def\uc{{\underline\cc}}           \def\ud{{\underline\dd}}
\def\up{{\underline\pp}}           \def\ut{{\underline t}}
\def\uxi{{\underline \xi}}         \def\umu{{\underline \m}}
\def\uv{{\underline\vv}}           \def\ue{{\underline \e}}
\def\uy{{\underline\yy}}           \def\uz{{\underline \zz}}
\def\uw{{\underline \ww}}          \def\uo{{\underline \o}}

\def\ha{{\widehat \a}}      \def\hx{{\widehat \x}}
\def\hb{{\widehat \b}}      \def\hr{{\widehat \r}}
\def\hw{{\widehat w}}       \def\hv{{\widehat v}}
\def\hf{{\widehat \f}}      \def\hW{{\widehat W}}
\def\hH{{\widehat H}}       \def\hB{{\widehat B}}
\def\hh{{\widehat \h}}      \def\hu{{\widehat u}}
\def\hK{{\widehat K}}       \def\hU{{\widehat U}}
\def\hp{{\widehat \ps}}     \def\hF{{\widehat F}}
\def\bp{{\bar \ps}}         \def\hc{{\hat \c}}
\def\jm{{\jmath}}           \def\hJ{{\widehat \jmath}}
\def\hJ{{\widehat J}}       \def\hg{{\widehat g}}
\def\tg{{\tilde g}}         \def\hQ{{\widehat Q}}
\def\hP{{\widehat P}}       \def\hC{{\widehat C}}
\def\hA{{\widehat A}}       \def\hD{{\widehat \D}}
\def\hDD{{\hat \D}}         \def\bl{{\bar \l}}
\def\hG{{\widehat G}}       \def\hS{{\widehat S}}
\def\hR{{\widehat R}}       \def\hM{{\widehat M}}
\def\hN{{\widehat N}}       \def\hn{{\widehat \n}}
\def\bk{{\bar \kk}}
\def\tv{{\tilde v}}

\def\tA{\tilde A}
\def\tit{\tilde t}
\def\tc{\tilde c}

\def\be#1\ee{\begin{equation}#1\end{equation}}
\def\bsp#1\esp{\begin{split}#1\end{split}}
\def\bal#1\eal{\begin{align}#1\end{align}}
\def\bald#1\eald{\begin{aligned}#1\end{aligned}}
\def\ba#1#2\ea{\begin{array}{#1}#2\end{array}}

\def\bea{\begin{eqnarray}}   \def\eea{\end{eqnarray}}
\def\bean{\begin{eqnarray*}} \def\eean{\end{eqnarray*}}
\def\bfr{\begin{flushright}} \def\efr{\end{flushright}}
\def\bc{\begin{center}}      \def\ec{\end{center}}
\def\bd{\begin{description}} \def\ed{\end{description}}
\def\bv{\begin{verbatim}}

\def\bra#1{{\langle#1|}}         \def\ket#1{{|#1\rangle}}
\def\lft{\left}                  \def\rgt{\right}
\def\la{{\langle}}               \def\ra{{\rangle}}
\def\ua{{\uparrow}}              \def\da{{\downarrow}}
\def\uda{{\uparrow\!\downarrow}}
\def\lp{{\hskip-1pt:\hskip 0pt}} \def\rp{{\hskip-1pt :\hskip1pt}}
\def\norm#1{{\left|\hskip-.05em\left|#1\right|\hskip-.05em\right|}}

\def\erp{{\perp}}
\def\arp{{\parallel}}

\let\mmp=\mp

\def\mp#1{\marginpar{\tiny\bf#1}}
\def\fb#1{{\bf#1}}

\def\nn{\nonumber}
\def\Halmos{\hfill\vrule height10pt width4pt depth2pt \par\hbox to \hsize{}}
\def\pref#1{(\ref{#1})}
\def\virg{\;,\quad}
\def\qed{\raise1pt\hbox{\vrule height5pt width5pt depth0pt}}
\let\dpr=\partial
\let\circa=\cong
\let\bs=\backslash
\let\==\equiv
\let\txt=\textstyle
\let\dis=\displaystyle
\let\io=\infty
\let\0=\noindent

\def\pagina{{\vfill\eject}}
\def\*{\vspace{.2cm}}
\def\ie{\hbox{\it i.e.\ }}
\def\eg{\hbox{\it e.g.\ }}
\def\tilde#1{{\widetilde #1}}
\def\der{\hbox{\rm d}}
\def\defi{{\buildrel \;def\; \over =}}
\def\apt{{\;\buildrel apt \over =}\;}
\def\nequiv{\not\equiv}
\def\Tr{\rm Tr}
\def\diam{{\rm diam}}
\def\sgn{\rm sgn}

\def\gtopl{\hbox{\msxtw \char63}}
\def\ltopg{\hbox{\msxtw \char55}}

%
\def\ins#1#2#3{\vbox to0pt{\kern-#2 \hbox{\kern#1 #3}\vss}\nointerlineskip}

\newdimen\xshift \newdimen\xwidth \newdimen\yshift
\newcount\griglia

\def\insertplot#1#2#3#4#5#6{%
\xwidth=#1pt \xshift=\hsize \advance\xshift by-\xwidth \divide\xshift by 2%
\begin{figure}[ht]
\vspace{#2pt} \hspace{\xshift}
\begin{minipage}{#1pt}
#3 \ifnum\driver=1 \griglia=#6
\ifnum\griglia=1 \openout13=griglia.ps \write13{gsave .2
setlinewidth} \write13{0 10 #1 {dup 0 moveto #2 lineto } for}
\write13{0 10 #2 {dup 0 exch moveto #1 exch lineto } for}
\write13{stroke} \write13{.5 setlinewidth} \write13{0 50 #1 {dup 0
moveto #2 lineto } for} \write13{0 50 #2 {dup 0 exch moveto #1
exch lineto } for} \write13{stroke grestore} \closeout13
\includegraphics{griglia.ps} \fi
\includegraphics{#4.ps}\fi%
\ifnum\driver=2 \fi
\end{minipage}
\caption{#5}
\end{figure}
}


\def\lb#1{\label{#1}}

\driver=1


\title{Weyl semimetallic phase in an interacting lattice system}

\author{Vieri Mastropietro}
\author{\vspace{5pt} Vieri Mastropietro\\
  \vspace{-4pt}\small{Dipartimento di Matematica, Universit\`a di Milano} \\
  \small{Via Saldini, 50, I-20133 Milano, ITALY }\\
  }



\maketitle
\begin{abstract}														
By using Wilsonian Renormalization Group (RG) methods
we rigorously establish the existence of a Weyl semimetallic phase in an interacting three dimensional fermionic lattice system,
by showing that the zero temperature Schwinger functions 
are asymptotically close to the ones of massless Dirac fermions. 
This is done via an expansion which is convergent in a region of parameters, which
includes the quantum critical point discriminating between the semimetallic and the insulating phase.
\end{abstract}

\section{Model and 
results}

\subsection{Introduction}

There is an increasing interest in materials whose 
Fermi surface is not extended, as it is usually the case, but it consists of disconnected points.
In several of such materials the charge carriers admit at low energies an effective description in terms 
of Dirac massless particles. This opens the exiting possibility that high energy phenomena have a counterpart
at low energies in real materials.
{\it Graphene} is probably the most known example of such systems;
it was pointed out in \cite{W1},\cite{S1}
that fermions on the honeycomb lattice 
behave as massless Dirac fermions in $2+1$ dimensions
and indeed the experimental realization of graphene \cite{GN}, a monolayer sheet of graphite,
offered a spectacular physical realization of such a system.  

As a next step, it is natural to look 
for materials with
electronic bands touching in couples of points and with an emerging description in terms
of $3+1$ massless Dirac (or Weyl) particles, the
 same appearing in the standard model to describe quarks and leptons. Such systems have been  called 
{\it Weyl semimetals}, and their existence has been predicted in several systems 
\cite{1},\cite{1a},\cite{1b},\cite{J},\cite{DC}; this has generated an intense experimental research,
see for instance \cite{36}, \cite{37} (and the review \cite{VV1}).
It is of course important to understand the effect of the interactions, which are usually analyzed
in  effective relativistic
models neglecting lattice effects \cite{40},\cite{41},
\cite{RL},\cite{MN}. Perturbative considerations suggest that short range interactions
can generate instabilities only at strong coupling, but in order to exclude non perturbative effects one has to prove the convergence of the expansions. It is also known that in such class of systems 
the effective relativistic description
misses important features; for instance in the case of graphene the universality of the optical conductivity emerges only taking into account the lattice \cite{GMPcond}. 

In this paper we consider
a  three dimensional interacting fermionic {\it lattice} model
with a point-like Fermi surface \cite{DC} (see also \cite{J}), in presence
of an Hubbard interaction. We construct the zero temperature correlations for couplings
not too large, proving the persistence of the Weyl semimetallic phase in presence of interactions.
In the non interacting case the semimetallic phase, in which  
the elementary excitations are well described in terms of Weyl fermions, is present in an extended region
of the parameters; outside such a region an insulating behavior is present
and a quantum critical point discriminates between the two phases.
In the semimetallic phase the Fermi surface consists of two points; close to the critical point the two points are very close and the Fermi velocity is arbitrarily small (and vanishes at the boundary).  
The effective relativistic description coincides with a system of massless Dirac fermions
in $3+1$ dimensions with an ultraviolet cut-off like the Gross-Neveu model or QED with massive photon:
in such models
the interaction is irrelevant and the convergence of the renormalized perturbative expansion has been established, see \cite{PPO} and \cite{M2}.
However the convergence radius in such models is {\it vanishing} with the particle velocity; therefore
such results give essentially no information for lattice Weyl semimetals close to the boundary of the semimetallic phase  where the Fermi velocity is very small. 
One may suspect that 
even an extremely  weak interaction could produce some quantum instability close to the boundary of the semimetallic phase, where the parameters correspond to a strong coupling regime in the effective description.
This is however excluded by the present paper: we can prove the persistence of the Weyl semimetallic phase
in presence of interaction in all the semimetallic region, even arbitrarily close to the boundary 
where the Fermi velocity vanishes. This result is achieved writing the correlations
in terms of a renormalized expansion with a radius of convergence which is independent from 
the Fermi velocity, and in order to get this  
one needs to exploit the non linear corrections to the dispersion relation due to the lattice. The proof is indeed based 
on two different multiscale analysis  in two regions of 
the energy momentum space; in the smaller energy region the effective relativistic description is valid while in the larger energy region
the quadratic corrections due to the lattice are dominating. In both regimes the interaction is irrelevant 
but the scaling dimensions are different; after the integration of the first regime one gets gain factors 
which compensate 
exactly the velocities at the denominator produced in the second regime, so that uniformity is achieved.
Such a phenomenon is completely absent in Graphene, in which the the Fermi velocity
is essentially constant. Another important phenomenon present here (and absent either in Graphene
and in the effective relativistic description) is 
the movement of Weyl points due to the interaction.

The analysis is based 
on the {\it Renormalization Group} (RG)  method of {\it Wilson}
and its approach to the effective action \cite{W}, in the form implemented in \cite{G} and \cite{Po}
in the context of perturbative renormalization. It was realized in the eighties that such methods
can be indeed used to get a
full {\it non-perturbative} control of certain fermionic Quantum Field Theories in $d=1+1$, \cite{GK}, \cite{L}
using Gram bounds and Brydges formula for truncated expectation \cite{B}. 
A very natural development was then to apply such techniques to condensed matter models,
\cite{BG},\cite{FMRT} with the final aim at obtaining a full non perturbative control of the ground state properties of interacting systems. However, while the interaction in the models considered in 
 \cite{GK} or \cite{L} is marginally irrelevant or dimensionally irrelevant, this is not the case
in interacting non relativistic fermionic models 
in one dimension, or in dimensions greater than one with extended Fermi surface.
This is due to the fact 
that the ground state properties of the interacting system are generically different with respect to the non interacting case. In one dimension 
it was finally obtained a full control of the zero temperature properties of interacting fermions, in the spinless \cite{BGPS},\cite{BM1} or repulsive spinning case \cite{BFM}; this was achieved
by combining
RG methods with Ward Identities based on the emerging chiral symmetries.
In systems in higher dimensions with extended symmetric Fermi surface, rigorous results 
were obtained, see 
\cite{DR}, \cite{BGM}, for temperatures {\it above} an 
exponentially small scale setting the onset of (possible) quantum instabilities.
Only in the case of an {\it asymmetric} Fermi surface (a condition preventing the formation
of Cooper pairs) the convergence of the renormalized expansion up to zero temperature for a interacting fermionic system was achieved 
\cite{FKT}, proving the existence of a Fermi liquid phase.
In
systems with point-like Fermi surfaces in two or three dimensions the interaction is irrelevant and this allows
the proof of the convergence of the renormalized expansion up to zero temperature, as in the case of Graphene \cite{GM}
or the case discussed in the present paper. In the case of Graphene, the combination of 
non perturbative bounds with lattice Ward Identities 
allows to establish remarkable physical conclusions, like the universality of the optical conductivity  \cite{GMPcond}.
Similarly, Ward Identities combined with the results obtained in the present paper
can be used to establish
a weak form of universality for the optical conductivity in Weyl semimetals, see 
\cite{M4}.


\subsection{The model}

We consider the interacting version of the
tight binding model introduced in 
\cite{DC}, describing fermions on a three dimensional lattice, 
with nearest and next to nearest neighbor hopping and with a properly defined magnetic flux density, whose effect is to decorate the hopping with phase factors \cite{Ha}. 

We consider two cubic sublattices $\L_A=\L$ and $\L_B$, where 
$\L_B=\L_a+\vec\d_+$ and $\L=\{n_1\vec\d_1+n_2\vec\d_2+n_3\vec\d_3, n_1,n_2,n_3=0,1,...,L-1\}$
with 
$\vec\d_1=(1,0,0)$, 
$\vec\d_2=(0,1,0)$, $\vec\d_3=(0,0,1)$ and $\vec\d_{\pm}={\vec\d_1\pm\vec\d_2\over 2}$. 
We introduce creation and annihilation
fermionic operators for electrons sitting at the sites of the A- and B- sublattices; if $\vec x\in\L$ 
\be
a^\pm_{\vec x}={1\over  |\L|}\sum_{\vec k\in\DD_L} e^{\pm i \vec k\vec x}\hat a^\pm_{\vec k}\quad 
b^\pm_{\vec x+\vec \d_+}={1\over  |\L|}\sum_{\vec k\in\DD_L} e^{\pm i \vec k\vec x}\hat b^\pm_{\vec k}
\ee
with $\DD_L=\{\vec k={2\pi\over L}\vec n$, $\vec n=(n_1,n_2,n_3)$, $n_i=(0,1,...,L-1)\}$
and
\be
\{\hat a^\e_{\vec k},\hat a^{-\e'}_{\vec k'} \}=|\L|\d_{\vec k,\vec k'}\d_{\e,\e'}\quad\quad \{\hat b^\e_{\vec k},\hat b^{-\e'}_{\vec k'} \}=|\L|\d_{\vec k,\vec k'}\d_{\e,\e'}
\ee 
and $\{\hat a^\e_{\vec k}, \hat b^{\e'}_{\vec k}\}=0$.

The hopping (or non-interacting) Hamiltonian is given by, if $\hat\psi^\pm_{\vec k}=(\hat a^\pm_{\vec k},\hat b^\pm_{\vec k})$ 
\be
H_0={1\over |\L|}\sum_{\vec k\in D_L}(\hat\psi^+_{\vec k},\EE(\vec k)\hat\psi^-_{\vec k})
\ee
where, if $k_\pm=\vec k\vec\d_\pm$
\be\EE(\vec k)=t\sin (k_+)\s_1+t\sin (k_-)\s_2+
\s_3(\m+t_\perp \cos k_3- {1\over 2}t'(\cos k_1+\cos k_2))
\ee
and
$$\s^1=
 \begin{pmatrix}&0&1\\
          &1&0\end{pmatrix}
 \quad
 \s^2=\begin{pmatrix}
 &0&-i\\ &i&0
\end{pmatrix}
 \quad\s^3=\begin{pmatrix}&1&0\\
          &0&-1\end{pmatrix}$$ 

%
%
The hopping parameters $t,t_\perp,t'$ are assumed $O(1)$ and positive; in coordinate space $t_\perp$
describes the hopping between fermions living in different horizontal layers, $t$
the nearest neighbor hopping in the same layer, $t'$ the next-to-nearest neighbor hopping, while $\m$
the difference of energy between $a$ and $b$ fermions. The hopping terms are multiplied by suitable phases 
to take into account a magnetic flux pattern applied to the lattice.

The electrons on the lattice can interact through a short range (or Hubbard) two body interaction, so that the total Hamiltonian is  
\be
H=H_0+V\label{ham}
\ee
where
\be
V=U\sum_{\vec x,\vec y} v(\vec x-\vec y) 
[a^+_{\vec x} a^-_{\vec x}+
b^+_{\vec x+\vec \d_+} b^-_{\vec x+\vec \d_+}][
 a^+_{\vec y} a^-_{\vec y}+ b^+_{\vec y+\vec \d_+} b^-_{\vec y+\vec \d_+}
]\ee
and $|v(\vec x)|\le C e^{-\k|\vec x|}$ is a short-range interaction ($C,\k$ positive constants).

Defining $\psi^\pm_{\vec x}=(a^\pm_{\xx},b^\pm_{\xx+\d_+})$,
we consider the operators
\be
\psi^\pm_{\xx}=e^{x_{0}H}\psi^\pm_{\vec
x}e^{-x_{0}H}\ee with $\xx=(x_{0},\vec x)$ and  $0<x_0<\b$
and $\b^{-1}$ is the temperature; on $x_0$ antiperiodic boundary conditions are imposed. 
The 2-point {\it Schwinger function} is defined as
\be
S_{U}(\xx-\yy)={\Tr\{e^{-\b H} \psi^{-}_{\xx}\psi^{+}_{\yy}
\}\over 
\Tr e^{-\b H}}\equiv \media{{\bf T}\{ \psi^{-}_{\xx_1}\psi^{+}_{\yy}\}
}_{\b,\L}
\label{sf}
\ee
where ${\bf T}$ is the fermionic time ordering operation.

\subsection{The non interacting case}

%
%
%
%
The Hamiltonian in the non interacting $U=0$ case can be easily written in diagonal form 
\be
H_0={1\over |\L|}\sum_{\vec k\in \DD_L}[\l(\vec k)
\hat \a^+_{\vec k} \a^-_{\vec k}-\l(\vec k)\hat\b^+_{\vec k}\b^-_{\vec k}]
\ee
where 
\bea
&&\l(\vec k)=\\
&&\sqrt{t^2(\sin^2 (k_+)+\sin^2 (k_-))+
(\m+t_\perp \cos k_3- {1\over 2}t'(\cos k_1+\cos k_2))^2}\nn
\eea
where $\hat\a^\pm_{\vec k},\hat \b^\pm_{\vec k}$ are sitable linear combinations of $\hat a^\pm_{\vec k},\hat b^\pm_{\vec k}$.
If $-\b<x_0-y_0\le \b$, the 2-point Schwinger function is given by
\bea &&\media{{\bf T}\{\a_{\xx}^-\a_{\yy}^+\}}_{\b,\L}=
{1\over|\L|}\sum_{\vec k\in\DD}e^{-i\vec k( \vec x-\vec
y)}\Big[\chi\big(x_0-y_0>0\big)\frac{e^{(x_0-y_0) \l(\vec
k)}} {1+e^{\b\l(\vec k)
}}-\nn
\\
&&\chi\big(x_0-y_0\le
0\big)
\frac{e^{(x_0-y_0+\b) \l(\vec k)}} {1+e^{\b\l(
\vec
k)}} \Big]
\label{A.19aa}
\eea
\bea
 &&\media{{\bf T}\{\b_{\xx}^-\b_{\yy}^+\}}_{\b,\L}=
{1\over|\L|}\sum_{\vec k\in\DD}e^{-i\vec k( \vec x-\vec
y)}\Big[\chi\big(x_0-y_0>0\big)\frac{e^{-(x_0-y_0) \l(\vec
k)}} {1+e^{-\b\l(
\vec k)}}-\nn\\
&&\chi\big(x_0-y_0\le
0\big)\frac{e^{-(x_0-y_0+\b) \l(\vec k)}} {1+e^{-\b\l(
\vec
k)}} \Big] \label{A.19}\eea
and $\media{{\bf T}\{\a_{\xx}^-\b_{\yy}^+\}}_{\b,\L}=
\media{{\bf T}\{\b_{\xx}^-\a_{\yy}^+\}}_{\b,\L}=0$. A priori 
Eq.(\ref{A.19aa}) and (\ref{A.19}) are defined only for
$-\b<x_0-y_0\le \b$, but we can extend them periodically over the 
whole real axis; the periodic extension of the propagator is 
continuous in the time variable for $x_0-y_0\not\in\b \ZZZ$, and it has
jump discontinuities at the points $x_0-y_0\in\b\ZZZ$. 
Note that at $x_0-y_0=\b n$, the difference between the right and left 
limits is equal to
$(-1)^n\d_{\vec x,\vec y}$, so that the propagator is discontinuous only
at $\xx-\yy=\b\ZZZ\times \vec 0$. For $\xx-\yy\not\in\b\ZZZ\times \vec 0$, 
we can write, defining $\DD=D_L\times \DD_\b$, $\DD_\b=\{k_0={2\pi\over\b}(n_0+{1\over 2}), n_0\in \ZZZ\}$,
$\kk=(k_0,k)$
\bea &&\media{{\bf T}\{\a_{\xx}^-\a_{\yy}^+\}}_{\b,\L}=
\frac{1}
{\b|\L|}\sum_{\kk\in\DD_{\b,L}}e^{-i\kk(\xx-\yy)}
\frac1{-ik_0-
\l(\vec k)}\;, \label{A.19a1}\\
&&\media{{\bf
T}\{\b_{\xx}^-\b_{\yy}^+\}}_{\b,\L}=\frac{1}
{\b|\L|}\sum_{\kk\in\DD_{\b,L}}e^{-i\kk(\xx-\yy)}
\frac1{-ik_0+ \l(\vec k)}\;. \label{A.19a2}\eea
 If we now re-express
$\a_{\xx}^\pm$ and $\b^\pm_{\xx,}$ in terms of
$a^\pm_{\xx,}$ and $b^\pm_{\xx+\dd_1,}$ we get
for $\xx-\yy\not\in\b\ZZZ\times \vec 0$
\be
S_0(\xx-\yy)={1\over |\L|\b}\sum_{\kk\in\DD_{L,\b}}
e^{i\kk(\xx-\yy)}A^{-1}(\kk)
\ee
where
\be
A(\kk)=-i k_0 I+t\s_1\sin k_++t\s_2 \sin k_-+(\m-t'+t_\perp \cos k_3+E(\vec k)
)\s_3
\ee
with
\be
E(\vec k)=t'(\cos k_+\cos k_--1)
\ee
The {\it Fermi surface} is defined as the singularity of the Fourier transform of the 2-point function $\hat S_0(\kk)=A^{-1}
(\kk)$ at zero temperature and $k_0=0$. Note that the functions $\sin (k_+)$ and $\sin (k_-)$ vanish in correspondence of two points 
$(k_1,k_2)=(0,0)$ and $(k_1,k_2)=(\pi,\pi)$ and we will assume from now on 
\be
\m+t'>2 t_{\perp}\label{a1}
\ee
so that
the only possible singularities are when 
$\m-t'+t_\perp \cos k_3=0$. Therefore if
\be
{|\m-t'|\over t_\perp}<1\label{a2}
\ee
than $\hat S_0(\kk)$ is singular in correspondence of two points, called
{\it Weyl points} and denoted by
$\pm \vec p_F$, with 
\be
\vec p_F=(0,0,\cos^{-1}({t'-\m\over t_\perp}))\label{w}
\ee
Close to such points
the 2-point function has the following form, if $\kk=\kk'\pm \pp_F$
and $|\kk'|<<|\sin p_F|$
\bea
\hat S_{0}(\kk'\pm \pp_F)\sim 
\begin{pmatrix}&-i k_0\pm v_{3,0} k'_{3}
& v_{0}(k_+-i k_-)\\
&
v_{0}(k_++i k_-)& 
-i k_0-(\pm) v_{3,0}  k'_3
\end{pmatrix}^{-1}\label{rel}
\eea
with
\be
v_{0}=t\quad\quad  v_{3,0}=t_\perp \sin p_F
\ee
The two $2\times 2$ matrices $\hat S_0(\kk'+\pp_F)$ and 
$\hat S_0(\kk'-\pp_F)$ can be combined in a $4\times 4$ matrix coinciding with the propagator of a massless
Dirac (or Weyl) particle in $D=3+1$ dimension, with an {\it anisotropic} light velocity. 
In coordinate space, the 2-point function has a power law
decay times an oscillating factor, denoting a metallic behavior (or semimetallic, as the conductivity computed via Kubo formula vanishes at zero frequency) under the conditions \pref{a1} and \pref{a2}.

On the contrary for ${|\m-t'|\over t_\perp}>1$
the 2-point function decays exponentially for large distances ($\hat S_0(\kk)$ is non singular)
and the system has an {\it insulating} behavior.
Close to the boundaries of the semimetallic phase, the Fermi velocity $t_{\perp}\sin p_F$ becomes arbitrarily small and the Weyl points are very close; the relativistic behavior \pref{rel} emerges only in a very small region 
$|\kk'|<<t_\perp \sin p_F$
around the Fermi points as
the linear dispersion relation $v_{3,0} k'_3$ is dominating over the quadratic correction only in that region.
There is a {\it quantum critical point}  ${|\m-t'|\over t_\perp}=1$ discriminating the metallic and the insulating region.

We ask now the question if Weyl semimetallic behavior, present under the conditions \pref{a1} and \pref{a2},
survives to the presence of the interaction.

\subsection{Grassmann Integral representation}

The analysis of the interacting case is done by a rigorous implementation of RG techniques. The starting point is a functional integral representation of the Schwinger functions which is quite suitable for such methods. 
We want to establish the persistence of Weyl semimetallic behavior with Weyl points given by 
\pref{w}. However, even if the semimetallic phase persists in presence of interaction, there is no reason {\it a priori} for which the value of $p_F$ should be the same in the free or interacting case.
Therefore it is convenient to proceed
in two steps. The first consists in writing
$\m=\m-\n+\n=\bar\m+\n$ and in proving that one can choose $\n=\n(\bar\m,\l)$ so that there is Weyl semimetallic behavior under the condition ${|\bar\m-t'|\over t_\perp}<1$, and that in such region the Weyl points are given by $(0,0,\pm p_F)$ with $\cos p_F={|\bar\m-t'|\over t_\perp}$;
in this way the location of the singularity of the two point function does not move, and this is technically convenient as we construct the interacting function as series starting from the non interacting one.
Once that this is 
(possibly) done the second step consists in solving the inversion problem $\bar\m+\n(\bar\m,\l)=\m$,
so that one can determine the location of the Weyl points as function  of the initial parameters. We will
deal here with the first step only, which is the substantial one; the inversion problem
can be done via standard methods once the first step is done, see for instance Lemma 2.8 of \cite{BFM}
for a similar problem. 

We will introduce a set of {\it Grassmann variables} 
$\hat\psi^\pm_{\kk}=(\hat a^\pm_\kk,\hat b_{\kk})$, $\kk\in \DD_{\b,L}$ 
by the same symbol as the fermionic fields. We also define a "regularized" propagator 
$g_M(\xx-\yy)$ ($2^{M}$ is an ultraviolet cut-off)
with 
\be
g_M(\xx-\yy)=
{1\over |\L|\b}\sum_{\kk\in\DD_{L,\b}}
e^{i\kk(\xx-\yy)}A^{-1}(\kk)\bar\chi(2^{-M}|k_0|)
\ee
with 
$\bar\chi(t):\RRR^+\to \RRR$ is a smooth compact support function equal to $1$ for $0<t<1$ and $=0$ for $t>2$. Note that, contrary to the function $S_0(\xx-\yy)$, the sum $\sum_{\kk\in\DD_{L,\b}}$
is restricted over a finite number of elements. Note also that for $\xx-\yy\not=(0,n\b)$, $n\in \ZZZ$ than
\be
\lim_{M\to\io} g_M(\xx-\yy)=S_0(\xx-\yy)\label{fff}
\ee 
The above equality is however not true for $\xx-\yy=(0,n\b)$; indeed the r.h.s. of \pref{fff} is discontinuous while the l.h.s. is equal to ${1\over 2}[S_0(0,0^+)+ S_0(0,0^-)]$.

We introduce the {\it generating functional}
\be
e^{\WW_M(\phi)}=\int P(d\psi)e^{\VV(\psi)+(\psi,\phi)}\label{gf}
\ee
where $P(d\psi)$ is the fermionic "measure"  with propagator 
$g_M(\xx-\yy)$ and
$\VV$ is the interaction given by
\be
\VV=(\n+\n_C) N+ V
\ee
where, if $\int d\xx= \int dx_0\sum_{\vec x}$
\bea
&&N=\int d\xx\psi^+_{\xx}\s_3
\psi^-_{\xx}\\
&&V=U 
\int d\xx d\yy 
 v(\xx-\yy)(\psi^+_{\xx}I\psi^-_{\xx})(\psi^+_{\yy} I\psi^-_{\yy})\nn
\eea
if $v(\xx)=\d(x_0) v(x)$. Moreover 
$(\psi,\phi)=\int d\xx [\psi^+_{\xx}\s_0\phi^-_{\xx}+\psi^-_{\xx}\s_0\phi^+_{\xx}]$
and $\n_C=U \hat v(0)[S_0(0,0^+) -S_0(0,0^-)]$. We define
\be
S_{2}(\xx-\yy)=\lim_{M\to\io} S_{M}(\xx-\yy)=\lim_{M\to\io}
{\partial^2 \WW_M\over\partial\phi^+_{\xx}\partial\phi^-_{\yy}}\Big|_0\label{sds}
\ee
It is easy to check order by order in perturbation theory that 
$\lim_{M\to\io} S_{M,U}(\xx-\yy)$ coincides in the $M\to\io $ limit with $S_{U}(\xx-\yy)$ by 
\pref{sf} with $\m$ replaced by $\m+\n$. Note indeed that both functions can be expressed in terms of
the same Feynman diagrams with propagator respectively $S_0(\xx-\yy)$ and $g_M(\xx-\yy)$. Therefore the equality is trivial except in the graphs containing a {\it tadpole}, involving a propagator computed at $(0,0)$;
the presence of the countertern $\n_C$ ensures than the equality, see \S 2.1 of \cite{BFM} for more details in a similar case.

One can prove more; if  $S_{M}(\xx-\yy)$ 
given by \pref{sds}
is analytic and bounded in $|U|\le U_0$ with $U_0$ independent of $\b,L$
and uniformly convergent as $M\to\io$, then
$\lim_{M\to\io}S_{M}(\xx-\yy)=S_{U}(\xx-\yy)
$ where $S_{U}(\xx-\yy)$ is given by \pref{sf}; 
the proof of this fact is rather standard (it is an application of 
Weierstarss theorem and of properties of analytic functions) and it will be not repeated here (see
 Lemma 1 of \cite{GM} or prop 2.1 of \cite{BFM} for an explicit proof in similar cases). This ensures that 
one can study directly the Grassmann integral \pref{sds} 
to construct the 
Schwinger function
\pref{sf}. 

\subsection{Main results}

Our main result is the following.
 
\begin{theorem}
Let us consider $S_2(\xx)$ given by \pref{sds}
with $
\m+t'>2 t_{\perp}\label{a1}
$
and ${|\m-t'|\over t_\perp}<{3\over 2}$.
There exists $U_0>0$, independent of $\b,L $, such that if $|U|\le U_0$, it is possible to find a $\n$, analytic in $U$,
such that $S_2(\xx)$ exists and is analytic uniformly in $\b,L$ as $\b\to\io,L\to\io$. Moreover the Fourier transform of $S_2(\xx)$ in the $\b\to\io, L\to\io $ limit, denoted by $\hat S_2(\kk)$, in the case
${|\m-t'|\over t_\perp}<1$
is singular only at 
$\pm \pp_F$, with $\cos p_F={|\m-t'|\over t_\perp}$, $v_{3,0}=t_\perp\sin p_F$ and close to the singularity,
\bea
\hat S_{2}(\kk'\pm \pp_F)=
{1\over Z}\begin{pmatrix}&-i k_0\pm v_{3} k'_{3}
& v (k_+-i k_-)\\
&
v (k_++i k_-)& 
-i k_0-(\pm) v_{3} k'_3
\end{pmatrix}^{-1}(1+R(\kk'))\label{rel1}
\eea
with $|R(\kk)|\le {|\kk'|\over v_{3,0}}$ and  \be Z=1+O(U),
\quad {v_3-v_{3,0}\over v_{3,0}}=O(U),\quad v=v_{0}+O(U)\;.\ee On the other hand for 
${|\m-t'|\over t_\perp}>1$ the 2-point function is bounded for any $\kk$.
\end{theorem}
\vskip.3cm
{\bf Remarks}
\begin{enumerate}
\item The above theorem establishes analyticity in $U$ for values of the parameters including
either the semimetallic and the insulating phase, and proves for the first time
the existence of a Weyl semimetallic phase in an interacting system
with short range interactions. The effect
of the interaction is to generically
 modify the location of the Weyl points (the counterterm $\n$ takes 
this into account) and to change the parameters of the emerging relativistic description, like 
the wave function renormalization and the "light"  velocity. 
\item Note that close to the boundary of the semimetallic phase the (third component) of the Fermi velocity $v_3$ is small, and vanishes
continuously at the  
{\it quantum critical point} ${|\m-t'|\over t_\perp}=1$ discriminating between insulating and semimetallic phase. The estimated radius of convergence is uniform in $v_3$; this is remarkable as small $v_3$ 
correspond to a strong coupling regime in the effective relativistic description.
The main idea in order to achieve that is to perform a different multiscale analyis in two regions of the energy space, discriminated by an energy scale measuring the distance from the critical point.
\item The Renormalization Group analysis performed here to prove the above theorem can be 
used to determine the large distance behavior of the current-current correlations.
As a consequence, in combination with  
Ward Identities, some universality properties of the optical conductivity in the semimetallic phase can be proved, see \cite{M4}.
\end{enumerate}

\section{Renormalization Group analysis: First regime}

We find convenient the introduction of a parameter measuring the distance from the boundary of the semimetallic phase; therefore 
we define
\be {\m-t'\over t_\perp}=-1+r\ee
with $|r|\le {1\over 2}$; the case $r={1\over 2}$ corresponds, in the non interacting case, to the semimetal with the highest velocity $v_3$, while at $r=0$ the Fermi velocity $v_{3}$ vanishes. 

The starting point of the analysis of \pref{sds}
is the decomposition of the propagator in the following way
\be
g_M(\xx-\yy)=g^{(\le 0)}(\xx-\yy)+g^{(> 0)}(\xx-\yy)
\ee
where 
\bea
&&\hat g^{(\le 0)}(\kk)=\bar\chi(\g^{-M}|k_0|)\chi_{0}(\kk)\hat A^{-1}(\kk)\\
&&\hat g^{(> 0)}(\kk)=\bar\chi(\g^{-M}|k_0|)(1-\chi_{0}(\kk))\hat A^{-1}(\kk)
\eea
and
$\chi_{<0}(\kk)=\bar\chi(a_0^{-1}|\det A(\kk)|^{1\over 2})$, with $a_0={t_\perp\over 10}$.
The above decomposition corresponds to a decomposition in the Grassmann variables
$\psi=\psi^{(\le 0)}+\psi^{(>0)}$ with propagators respectively 
$g^{(\le 0)}(\xx)$  (the {\it infrared} propagator)
and $g^{(> 0)}(\xx)$ (the {\it ultraviolet} propagator).
We can write 
\bea
&&e^{\WW(\phi)}=\int P(d\psi^{(> 0)}) P(d\psi^{(\le 0)})
e^{\VV(\psi^{(> 0)}+\psi^{(\le 0)}
)+(\psi^{(> 0)}+\psi^{(\le 0)}
,\phi)}=\nn\\
&&=e^{\b |\L| E_0}
\int P(d\psi^{(\le 0)})
e^{\VV^{(0)}(\psi^{(\le 0},\phi)
}\label{ss}
\eea
with 
\be \VV^{(0)}(\psi,\phi)=\sum_{n,m\ge 0}
\int d\underline\xx \int d\underline\yy
[\prod_{i=1}^n\psi^{\e_i}_{\xx_i}][\prod_{i=1}^m\phi^{\e_i}_{\xx_i}]
W_{n,m}(\underline\xx,\underline\yy)\label{irr}
\ee
with $[\prod_{i=1}^n\psi^{\e_i}_{\xx_i}]=1$ if $n=0$ and $[\prod_{i=1}^m\phi^{\e_i}_{\yy_i}]=1$ if $m=0$, and for $U,\n$ smaller than a constant (independent from $L,\b,M$)
\be
{1\over \b |\L|}\int d\underline\xx \underline d\yy
|W^{(0)}_{n,m}(\underline\xx,\underline\yy)|\le \b |\L| C^{n+m}|U|^{max[1,n-1]}
\ee
Moreover $\lim_{M\to\io}W^{(0)}_{n,m}(\underline\xx,\underline\yy)$ and is reached uniformly.
The above properties follow from  
Lemma 2 of \cite{BFM} (app. B) or Lemma 2.2 of \cite{BFM}; the proofs in such papers 
are written for $d=2$ or $d=1$ lattice system, but the adaptation to the present case is straightforward 
(due to the presence of a spatial lattice the ultraviolet problem 
is essentially independent from dimension).

%
%
The infrared negative scales are
divided in two different regimes, which have to be analyzed 
differently as they have different scaling properties. 
They are discriminated by a scale \be h^*=[min(\log_2 a_0^{-1}10 |r|,0)]\ee 
which discriminates the region where the non-linear corrections to the dispersion relation are dominating with the region where the energy is essentially linear; if $h^*=0$ the first regime described here
is absent. We describe now the integration of the scales $h\ge h^*$ inductively. Assume that we have integrated already 
the scales $0,-1,..,h+1$
showing that \pref{gf} can be written as (in the $\phi=0$ for definiteness)
\be
e^{|\L| \b E_h}\int  P(d\psi^{(\le h)}) e^{\VV^{(h)}(\sqrt{Z_{h}}\psi^{(\le h)})}\label{ef1}
\ee
where $P(d\psi^{(\le h)})$ has propagator given by
%
\be
g^{(\le h)}(\xx)=\int d\kk e^{i\kk\xx}{\chi_h(\kk)\over Z_h} A_h^{-1}(\kk)
\ee
where
\bea
&&A_h(\kk)=\\
&&\begin{pmatrix}&-i k_0+v_{3,h}(\cos k_3-1+r+E(\vec k)& 
v_{h}(\sin k_+-i\sin k_-)\\
&
v_{h}(\sin k_++i \sin k_-)& 
-i k_0-v_{3,h}(\cos k_3-1+r-E(\vec k)
\end{pmatrix}\label{cond2}\nn
\eea
%
$
\chi_h(\kk)=\bar\chi(a_0^{-1} 2^{-h}|\det A_h(\kk)|^{1\over 2})$ and
\bea 
&&\VV^{(h)}(\psi)=\sum_{n\ge 1}
\int d\xx_1...\int d\xx_n 
\prod_{i=1}^n\psi^{\e_i}_{\xx_i}
W^{(h)}_{n}(\underline\xx)=\nn\\
&&{1\over (|\L|\b)^n}\sum_{\kk_1,...,\kk_n}\prod_{i=1}^n\hat\psi^{\e_i}_{\kk_i}
\hat W^{(h)}_{n}(\kk_1,..,\kk_{n-1})\d(\sum_{i=1}^n\e_i \kk_i)\nn
\eea
We introduce a {\it localization operator} acting on the effective potential as
\be \VV^{(h)}=\tilde \LL \VV^{(h)}+\RR \VV^{(h)}\label{loc} \ee
with $\RR=1-\tilde \LL$ and $\tilde\LL$ is a linear operator acting on the kernels 
$\hat W^{(h)}_{n}(\kk_1,..,\kk_{n-1})$ in the following way:
\begin{enumerate}
\item $\tilde\LL \hat W^{(h)}_{n}(\kk_1,..,\kk_{n-1})=0$ if $n>2$.
\item If $n=2$
\bea
&&\tilde\LL  \hat W^{(h)}_{2}(\kk)= \\
&&\hat W^{(h)}_{2}(0)+k_0  
\hat W^{(h)}_{2}(0)+\sum_{i=+,-,3} \sin k_i  
\partial_i \hat W^{(h)}_{2}(0)+(\cos k_3-1)\partial^2_3 \hat W^{(h)}_{2}(0)\nn
\eea
\end{enumerate}
The definition of $\tilde\LL$ is written in the $L=\b=\io$ limit for definiteness but its expression
for $L,\b$ finite is straightforward.
By symmetry
\bea
&&\hat W^{(h)}_{2}(0)=\s_3 n_h\quad \partial_+ \hat W^{(h)}_{2}(0)=\s_1 b_{+,h}\quad
\partial_- \hat W^{(h)}_{2}(0)=\s_2 b_{-,h}
\nn\\
&& \partial_3 \hat W^{(h)}_{2}(0)=0\quad
\quad \partial_3^2 \hat W^{(h)}_{2}(0)=\s_3 b_{3,h}
\eea
Note also that, by definition $\LL\RR=0$.
We can include the local part of the effective potential in the fermionic integration, so
that \pref{ef1} can be rewritten as 
\be
e^{|\L| \b \tilde E_h}\int \tilde P(d\psi^{(\le h)}) e^{\LL\VV^{(h)}(\sqrt{Z_{h-1}}\psi^{(\le h)}
+\RR \VV^{(h)}(\sqrt{Z_{h-1}}\psi^{(\le h)}
)}\label{ef2}
\ee
where 
\be
\LL\VV^{(h)}(\psi^{(\le h)})=2^h \n_h \int d\xx\psi^{+(\le h)}_\xx\s_3\psi^{-(\le h)}_\xx
\ee
and $\tilde P(d\psi^{(\le h)})$ is the Grassmann integration with propagator
similar to \pref{cond2} with $Z_{h-1}(\kk),v_{h-1}(\kk),v_{3,h-1}(\kk)$ replacing 
 $Z_{h},v_{h},v_{3,h}$,  where 
\bea
&&
Z_{h-1}(\kk)=Z_h[1+\chi_h^{-1}(\kk) b_{0,h}]\nn\\
&&v_{h-1}(\kk)={Z_h\over Z_{h-1}(\kk)}
[v_h+\chi_h^{-1}(\kk) b_{+,h}]\label{bbe}\\
&&
v_{3, h-1}(\kk)={Z_h\over Z_{h-1}(\kk)}[v_{3,h}+\chi_h^{-1}(\kk) b_{3,h}]\nn
\eea
Now we write  $\tilde P(d\psi^{(\le h)})=P(d\psi^{(\le h-1)}) P(d\psi^{(h)})$ where  $P(d\psi^{(h)})$
has propagator similar to $\hat g^{(\le h)}$ \pref{cond2} with the following differences: a) $A_h$ is replaced by $A_{h-1}$, where 
$Z_{h-1}\equiv Z_{h-1}(0)$, $v_{h-1}\equiv v_{h-1}(0)$, $v_{3,h-1}=v_{3, h-1}(0)$; b) $\chi_h$ is replaced by $f_h$, a smooth compact support function with support in $c_1 2^{h-1}\le |\det A_{h-1}(\kk)|^{1\over 2}\le c_2 2^{h+1}$, with $c_1<c_2$
positive constants.

Assuming that $Z_h,v_{3,h},v_{\pm, h}$ are close $O(U)$ to their value at $h=0$, one has
for any $N$ the following bound
\be
|g^{(h)}(\xx)|\le {1\over Z_h}
 {2^{5h\over 2}\over 1+[2^h (|x_0|+|x_+|+|x_-|)+2^{h\over 2}|x_3|]^N}
\label{b}
\ee
where we have used that $a_0\g^h-|r|\le a_0\g^{h}(1-{1\over 10})$.
Therefore $k_0,k_\pm= O(2^h)$, $k_3=O(2^{h\over 2})$ for large negative $h$ and the bound \pref{b} follows by integration by parts.
Finally we perform the integration over $\psi^{(h)}$ obtaining
%

\be
e^{\b |\L|\tilde e_h+\VV^{(h-1}(\sqrt{Z_{h-1}}\psi^{(\le (h-1)})}=\int P(d\psi^{(h)}) e^{\LL\VV^{(h)}(\sqrt{Z_{h-1}}\psi^{(\le h)})+\RR\VV^{(h)}(\sqrt{Z_{h-1}}\psi^{(\le h)})}
\label{2.40cc}
\ee

obtaining an expression identical to \pref{ef1} with $h-1$ replacing $h$, $E_{h-1}=\tilde E_h+\bar
e_h$,
so that the procedure can be iterated. 
%
\subsection{Tree expansion for the effective potentials.}
The effective potential 
$\VV^{(h)}(\psi^{(\le h)})$ can be written in terms of a tree expansion, defined as
follows.


\insertplot{300}{150}{ \ins{30pt}{85pt}{$r$} \ins{50pt}{85pt}{$v_0$}
\ins{130pt}{100pt}{$v$} \ins{35pt}{-5pt}{$h$} \ins{52pt}{-5pt}{$h+1$}
\ins{135pt}{-5pt}{$h_{v}$} \ins{215pt}{-5pt}{$-1$} \ins{235pt}{-5pt}{0}
\ins{255pt}{-5pt}{$1$}} {fig50} {A renormalized tree for
$\VV^{(h)}$\lb{h2a}}{0}

\0 1) Let us consider the family of all trees which can be constructed
by joining a point $r$, the {\it root}, with an ordered set of $n\ge 1$
points, the {\it endpoints} of the {\it unlabeled tree},
so that $r$ is not a branching point. $n$ will be called the
{\it order} of the unlabeled tree and the branching points will be called
the {\it non trivial vertices}.
The unlabeled trees are partially ordered from the root
to the endpoints in the natural way; we shall use the symbol $<$
to denote the partial order.
Two unlabeled trees are identified if they can be superposed by a suitable
continuous deformation, so that the endpoints with the same index coincide.
It is then easy to see that the number of unlabeled trees with $n$ end-points
is bounded by $4^n$.
We shall also consider the {\it labeled trees} (to be called
simply trees in the following); they are defined by associating
some labels with the unlabeled trees, as explained in the
following items.

\0 2) We associate a label $h\le -1$ with the root and we denote
$\TT_{h,n}$ the corresponding set of labeled trees with $n$
endpoints. Moreover, we introduce a family of vertical lines,
labeled by an integer taking values in $[h,1]$, and we represent
any tree $\t\in\TT_{h,n}$ so that, if $v$ is an endpoint or a non
trivial vertex, it is contained in a vertical line with index
$h_v>h$, to be called the {\it scale} of $v$, while the root $r$ is on
the line with index $h$.
In general, the tree will intersect the vertical lines in set of
points different from the root, the endpoints and the branching
points; these points will be called {\it trivial vertices}.
The set of the {\it
vertices} will be the union of the endpoints, of the trivial
vertices and of the non trivial vertices; note that the root is not a vertex.
Every vertex $v$ of a
tree will be associated to its scale label $h_v$, defined, as
above, as the label of the vertical line whom $v$ belongs to. Note
that, if $v_1$ and $v_2$ are two vertices and $v_1<v_2$, then
$h_{v_1}<h_{v_2}$.

\0 3) There is only one vertex immediately following
the root, which will be denoted $v_0$ and cannot be an endpoint;
its scale is $h+1$.

\0 4) Given a vertex $v$ of $\t\in\TT_{h,n}$ that is not an endpoint,
we can consider the subtrees of $\t$ with root $v$, which correspond to the
connected components of the restriction of
$\t$ to the vertices $w\ge v$. If a subtree with root $v$ contains only
$v$ and an endpoint on scale $h_v+1$,
it will be called a {\it trivial subtree}.

\0 5) With each endpoint $v$ we associate one of the monomials
contributing to $\RR {\cal
V}^{(0)}(\psi^{(\le h_v-1)})$, corresponding to the terms 
in the r.h.s. of (\ref{irr}) (with $\psi^{(\le 0)}$
replaced by $\psi^{(\le h_v-1)}$) and a set $\xx_v$ of space-time
points (the corresponding integration variables in the $\xx$-space
representation); or a term corresponding to $\LL\VV^{(h_v-1)}(\psi^{(\le h_v-1)})$.

\0 6) We introduce a {\it field label} $f$ to distinguish the field variables
appearing in the terms associated with the endpoints described in item 5);
the set of field labels associated with the endpoint $v$ will be called $I_v$;
note that $|I_v|$ is the order of the monomial contributing
to $\RR {\cal V}^{(0)}(\psi^{(\le h_v-1)})$  or $\LL\VV^{(h_v-1)}(\psi^{(\le h_v-1)})$
and associated to $v$.

Analogously, if $v$ is not an endpoint, we shall
call $I_v$ the set of field labels associated with the endpoints following
the vertex $v$; $\xx(f)$ will denote the
space-time point of the Grassmann field variable with label $f$.

In terms of these trees, the effective potential ${\cal V}^{(h)}$, $h\le -1$,
can be written as
\be {\cal V}^{(h)}(\psi^{(\le h)}) + \b|\L| \lis e_{k+1}=
\sum_{n=1}^\io\sum_{\t\in\TT_{h,n}}
{\cal V}^{(h)}(\t,\psi^{(\le h)})\;,\label{2.41}\ee
where, if $v_0$ is the first vertex of $\t$
and $\t_1,\ldots,\t_s$ ($s=s_{v_0}$)
are the subtrees of $\t$ with root $v_0$,
${\cal V}^{(h)}(\t,\psi^{(\le h)})$
is defined inductively as follows:\\

i) if $s>1$, then
\be {\cal V}^{(h)}(\t,\psi^{(\le h)})={(-1)^{s+1}\over s!} \EE^T_{h+1}
\big[\bar{\cal V}^{(h+1)}(\t_1,\psi^{(\le h+1)});\ldots; \bar{\cal V}^{(h+1)}
(\t_{s},\psi^{(\le h+1)})\big]\;,\label{2.42}\ee
where $\EE^T_{h+1}$ denotes the {\it truncated expectation} with propagator $g^{(h)}$ and 
 
$\bar{\cal V}^{(h+1)}(\t_i,\psi^{(\le h+1)})$ is equal to $\RR{\cal
V}^{(h+1)}(\t_i,\psi^{(\le h+1)})$ if the subtree $\t_i$ contains
more than one end-point, or if it contains one end-point but it is
not a trivial subtree;
it is equal to $\RR{\cal V}^{(0)}(\psi^{(\le h+1)})$
or $\LL\VV^{(h+1)}(\psi^{(\le h+1)})$
if $\t_i$ is a trivial subtree;\\

ii) if $s=1$, then ${\bar V}^{(h+1)}(\t,\psi^{(\le h)})$ is equal to
$\big[\RR{\cal V}^{(h+1)}(\t_1,\psi^{(\le h+1)})\big]$
if $\t_1$ is not a trivial
subtree; it is equal to $\RR {\cal
V}^{(0)}(\psi^{(\le h+1)})- \RR{\cal V}^{(0)}(\psi^{(\le h)})\big]$ 
if $\t_1$ is a trivial subtree (and therefore its end-point $v$ has scale $h_v=h+2$).
\vskip.4cm

Using its inductive definition, the right hand side of (\ref{2.41}) can be
further expanded, and in order to describe the resulting expansion we need some
more definitions. We associate with any vertex $v$ of the tree a subset $P_v$ of $I_v$,
the {\it external fields} of $v$. These subsets must satisfy various
constraints. First of all, if $v$ is not an endpoint and $v_1,\ldots,v_{s_v}$
are the $s_v\ge 1$ vertices immediately following it, then
$P_v \subseteq \cup_i
P_{v_i}$; if $v$ is an endpoint, $P_v=I_v$.
If $v$ is not an endpoint, we shall denote by $Q_{v_i}$ the
intersection of $P_v$ and $P_{v_i}$; this definition implies that $P_v=\cup_i
Q_{v_i}$. The union ${\cal I}_v$ of the subsets $P_{v_i}\setminus Q_{v_i}$
is, by definition, the set of the {\it internal fields} of $v$,
and is non empty if $s_v>1$.
Given $\t\in\TT_{h,n}$, there are many possible choices of the
subsets $P_v$, $v\in\t$, compatible with all the constraints. We
shall denote ${\cal P}_\t$ the family of all these choices and ${\bf P}$
the elements of ${\cal P}_\t$.

With these definitions, we can rewrite
${\cal V}^{(h)}(\t,\psi^{(\le h)})$
in the r.h.s. of (\ref{2.41}) as:
\bea &&{\cal V}^{(h)}(\t,\psi^{(\le
h)})=\sum_{{\bf P}\in{\cal P}_\t}
{\cal V}^{(h)}(\t,\PP)\;,\nn\\
&&{\cal V}^{(h)}(\t,\PP)=\int d\xx_{v_0}
\widetilde\psi^{(\le h)}(P_{v_0})
K_{\t,\PP}^{(h+1)}(\xx_{v_0})\;,\label{2.43}\eea
where
\be \widetilde\psi^{(\le h)}
(P_{v})=\prod_{f\in P_v}\psi^{ (\le
h)\e(f)}_{\xx(f)}\label{2.44}\ee
and $K_{\t,\PP}^{(h+1)}(\xx_{v_0})$ is defined inductively by
the equation, valid for any $v\in\t$ which is not an endpoint,
\be K_{\t,\PP}^{(h_v)}(\xx_v)={1\over s_v !}
\prod_{i=1}^{s_v} [K^{(h_v+1)}_{v_i}(\xx_{v_i})]\; \;\EE^T_{h_v}[
\widetilde\psi^{(h_v)}(P_{v_1}\setminus Q_{v_1}),\ldots,
\widetilde\psi^{(h_v)}(P_{v_{s_v}}\setminus
Q_{v_{s_v}})]\;,\label{2.45}\ee
where $\widetilde\psi^{(h_v)}(P_{v_i}\setminus Q_{v_i})$ has a definition
similar to (\ref{2.44}). Moreover, if $v_i$ is an endpoint
$K^{(h_v+1)}_{v_i}(\xx_{v_i})$ is equal to one
of the kernels of the monomials contributing to
$\RR{\cal V}^{(0)}(\psi^{(\le h_v)})$ or 
$\LL{\cal V}^{(h_v)}(\sqrt{Z_{h_v}}\psi^{(\le h_v)})$; if $v_i$ is not an
endpoint, $K_{v_i}^{(h_v+1)}=K_{\t_i,\PP_i}^{(h_v+1)}$, 
where ${\bf P}_i=\{P_w, w\in\t_i\}$.

We further decompose ${\cal V}^{(h)}(\t,\PP)$, by using the
following representation of the truncated expectation in the r.h.s. of
(\ref{2.45}). Let us put $s=s_v$, $P_i\=P_{v_i}\setminus Q_{v_i}$;
moreover we order in an arbitrary way the sets $P_i^\pm\=\{f\in
P_i,\e(f)=\pm\}$, we call $f_{ij}^\pm$ their elements and we
define $\xx^{(i)}=\cup_{f\in P_i^-}\xx(f)$, $\yy^{(i)}=\cup_{f\in
P_i^+}\xx(f)$, $\xx_{ij}=\xx(f^-_{ij})$,
$\yy_{ij}=\xx(f^+_{ij})$. Note that $\sum_{i=1}^s
|P_i^-|=\sum_{i=1}^s |P_i^+|\=n$, otherwise the truncated
expectation vanishes.

Then, we use the {\it Brydges-Battle-Federbush} formula \cite{L} saying that, up to a sign, if $s>1$,
\be \EE^T_{h}(\widetilde\psi^{(h)}(P_1),\ldots,
\widetilde\psi^{(h)}(P_s))=\sum_{T}\prod_{l\in T}
g^{(h)}(\xx_l-\yy_l)
\int dP_{T}({\bf t})\; {\rm det}\, G^{h,T}({\bf t})\;,\label{2.46}\ee
where $T$ is a set of lines forming an {\it anchored tree graph} between
the clusters of points $\xx^{(i)}\cup\yy^{(i)}$, that is $T$ is a
set of lines, which becomes a tree graph if one identifies all the
points in the same cluster. Moreover ${\bf t}=\{t_{ii'}\in [0,1],
1\le i,i' \le s\}$, $dP_{T}({\bf t})$ is a probability measure with
support on a set of ${\bf t}$ such that $t_{ii'}={\bf u}_i\cdot{\bf u}_{i'}$
for some family of vectors ${\bf u}_i\in \RRR^s$ of unit norm. Finally
$G^{h,T}({\bf t})$ is a $(n-s+1)\times (n-s+1)$ matrix, whose elements
are given by
\be G^{h,T}_{ij,i'j'}=t_{ii'}g^{(h)}(\xx_{ij}-\yy_{i'j'})\;,
\label{2.48}\ee
with $(f^-_{ij}, f^+_{i'j'})$ not belonging to $T$.
In the following we shall use (\ref{2.44}) even for $s=1$, when $T$
is empty, by interpreting the r.h.s. as equal to $1$, if
$|P_1|=0$, otherwise as equal to ${\rm det}\,G^{h}=
\EE^T_{h}(\widetilde\psi^{(h)}(P_1))$. 
It is crucial to note that $G^{h,T}$ is a Gram matrix, i.e.,
the matrix elements in (\ref{2.48}) can be written in terms of scalar products, and therefore it can be bounded by the Gram-Hadamard inequality.

%

If we apply the expansion (\ref{2.46}) in each vertex of
$\t$ different from the endpoints, we get an expression of the form
\be {\cal V}^{(h)}(\t,\PP) = \sum_{T\in {\bf T}} \int d\xx_{v_0}
\widetilde\psi^{(\le h)}(P_{v_0}) W_{\t,\PP,T}^{(h)}(\xx_{v_0})
\;,\label{2.49}\ee
where ${\bf T}$ is a special family of graphs on the set of points
$\xx_{v_0}$, obtained by putting together an anchored tree graph
$T_v$ for each non trivial vertex $v$. Note that any graph $T\in
{\bf T}$ becomes a tree graph on $\xx_{v_0}$, if one identifies
all the points in the sets $\xx_v$, with $v$ an endpoint.
Given $\t\in\TT_{h,n}$ and the labels $\PP,T$,
calling $I_R$ the endpoints of $\t$ 
to which is associated $\RR\VV^{(0)}$ and 
$I_\n$ the end-points associated to $\LL\VV^{(h_v-1)}$.
, the explicit representation of $W_{\t,\PP,T}^{(h)}
(\xx_{v_0})$ in (\ref{2.49}) is
\bea &&
W_{\t,\PP, T}^{(h)}(\xx_{v_0}) =
\left[\prod_{v\in I_R}K_{v}^{(0)} (\xx_{v})\right]\prod_{v\in I_v} 2^{h_v}\n_{h_v}
 \;\cdot\label{2.50aaa}\\ &&\cdot\;
\Bigg\{\prod_{v\,\atop\hbox{\ottorm not e.p.}}{1\over s_v!} \int
dP_{T_v}({\bf t}_v)\;{\rm det}\,\tilde G^{h_v,T_v}({\bf t}_v)\Biggl[
\prod_{l\in T_v} 
\big[(\xx_l-\yy_l)^{\a_l}\partial^{\b_l}g^{(h_v)}(\xx_l-\yy_l)\big]\,\Biggr]
\Bigg\}\;,\nn\eea
where $K_{v}^{(0)} (\xx_{v})$ are the kernels of $\RR\VV^{(0)}$; the factors
$(\xx_l-\yy_l)^\a$ and the derivatives $\partial^\b$ in the above expression are produced by the $\RR$ operation and finally $\tilde G^{h_v,T_v}$ differs
from $G^{h_v,T_v}$ for the presence of extra derivatives due to the $\RR$ operation
(see \S 3 of \cite{BM} for more details in a  similar case). 
The functions 
appearing in the r.h.s. of \pref{bbe}, namely
$b_{0,h}$, $b_{\pm,h}$, $b_{3,h}$ can be written as derivatives of
\be\bar W^{(h)}_l=\sum_{n=2}^\io\sum_{\t\in\bar \TT_{n,h}} \sum_{{\bf P}\in{\cal P}_\t}\sum_{T\in {\bf T}}{1\over |\L|\b}\int d\xx_{v_0} W_{\t,\PP, T}(\xx_{v_0})\label{bar}
\ee
where $\bar\TT_{n,h}$ is the subset of $\TT_{h,h}$ such that a)$h_{v^*}=h+1$
where $v^*$ is the first non trivial vertex; b)there is at least an end-point associated to $\VV^{(0)}$.
Condition a) is due to the fact that, by construction, $\LL\RR=0$; condition b) is due to the fact that
the contributions with only $\n$-vertices are vanishing, as it can be easily verified by an explicit computation in momentum space (they are chain graphs and $\hat g^{(k)}(0)=0$).

The next goal is the proof of the following result.\\

\begin{lemma}
There exists a constant $\e_0$ independent of 
of $\b$, $L$ and $r$, such that for $|U|\le \e_0$
and $\max_{k\ge h}[|\n_k|, |Z_k-1|,|v_{k,i}-v_{0,i}]\le \e_0$, $i=\pm,3$ then for $h\ge h^*$
\be \frac1{\b |\L|}\int d\xx_1\cdots d\xx_{l}|W^{(h)}_{l}
(\xx_1,\ldots,\xx_{l})|\le
2^{h (7/2-5 l/4)} \,(C \e_0)^{max(1,l/2-1)}\;.\label{2.52sz}\ee
Moreover, if $\bar W^{(h)}_{l}$ is given  by
\pref{bar} 
\be \frac1{\b |\L|}\int d\xx_1\cdots d\xx_{l}|\bar W^{(h)}_{l}
(\xx_1,\ldots,\xx_{l})|\le
2^{h (7/2-5 l/4)}2^{{1\over 8} h} \,(C \e_0)^{max(1,l/2-1)}\;.\label{2.52sz1}\ee
with $C$ a suitable constant.
\end{lemma}
\vskip.3cm
{\it Proof.} 
Using the tree expansion described above
we find that the l.h.s. of (\ref{2.50aaa}) can be bounded from above
by
\bea && \sum_{n\ge 1}\sum_{\t\in {\cal T}_{h,n}}
\sum_{\PP\in{\cal P}_\t}\sum_{T\in{\bf T}}
\int\prod_{l\in T^*}
d(\xx_l-\yy_l) \left[\prod_{v\in I_R}K_{v}^{(0)} (\xx_{v})\right]\prod_{v\in I_v} 2^{h_v}|\n_{h_v}|
\cdot
\label{2.53}\\
&&\cdot\Bigg[\prod_{v\ {\rm not}\ {\rm e.p.}}{1\over s_v!}
\max_{{\bf t}_v}\big|{\rm det}\, G^{h_v,T_v}({\bf t}_v)\big|
\prod_{l\in T_v}
\big[|\xx_l-\yy_l|^{\a_l}|\partial^{\b_l}g^{(h_v)}(\xx_l-\yy_l)|\big]\,\Biggr]
\nn\eea
where $T^*$ is a tree graph
obtained from $T=\cup_vT_v$, by adding
in a suitable (obvious) way, for each endpoint $v_i^*$,
$i=1,\ldots,n$, one or more lines connecting the space-time points
belonging to $\xx_{v_i^*}$.

A standard application of Gram--Hadamard inequality, combined with
the dimensional bound on $g^{(h)}(\xx)$ given by (\ref{b}), implies that
\be |{\rm det} G^{h_v,T_v}({\bf t}_v)| \le
c^{\sum_{i=1}^{s_v}|P_{v_i}|-|P_v|-2(s_v-1)}\cdot\;
2^{{5\over 4}{h_v}
\left(\sum_{i=1}^{s_v}|P_{v_i}|-|P_v|-2(s_v-1)\right)}\;.\label{2.54}\ee
By the decay properties of $g^{(h)}(\xx)$ given by \pref{b}, it
also follows that
\be \prod_{v\ {\rm not}\ {\rm e.p.}}
{1\over s_v!}\int \prod_{l\in T_v} d(\xx_l-\yy_l)\,
||g^{(h_v)}(\xx_l-\yy_l)||\le c^n \prod_{v\ {\rm not}\ {\rm e.p.}}
{1\over s_v!} 2^{-h_v(s_v-1)}\;.\label{2.55}\ee
The bound on the kernels produced by the ultraviolet integration
implies that
\bea &&\int\prod_{l\in T^*\setminus\cup_v T_v}d(\xx_l-\yy_l)
\left[\prod_{v\in I_R}K_{v}^{(0)} (\xx_{v})\right]\prod_{v\in I_v} 2^{h_v}|\n_{h_v}|\nn\\
&&\le C^n \e_0^n \Big[\prod_{v\
{\rm e.p.}, |I_v|=2} 2^{h_{v'}(1+\d_v)} \Big] \;,\label{2.56}\eea
where $\d_v={1\over 2}$ if $v\in I_R$ and $|I_v|=2$
and $\d_v=0$ otherwise
;  the factors 
$\g^{(1+\d_v)h_{v'}}$ are due to fact that $\RR$ acts on the terms with $|I_v|=2$. 
Therefore 
the l.h.s. of (\ref{2.52sz}) can be bounded from above
by
\bea &&\sum_{n\ge 1}\sum_{\t\in {\cal T}_{h,n}}
\sum_{\PP\in{\cal P}_\t\atop |P_{v_0}|=l}\sum_{T\in{\bf T}}
C^n \e_0^n \Big[\prod_{v\ {\rm not}\ {\rm e.p.}} \frac{1}{s_v!}
2^{{h_v}\left(\sum_{i=1}^{s_v}{5|P_{v_i}|\over 4}-{5|P_v|\over 4}-{7\over 2}(s_v-1)\right)}\Big]]\cdot\nn\\
&&\Big[\prod_{v\ {\rm not}\ {\rm e.p.}}
2^{-(h_v-h_{v'})z(P_v)}\Big] \Big[\prod_{v\
{\rm e.p.}, |I_v|=2} 2^{h_{v'}(1+\d_v)} \Big] \label{2.57aab}\eea
where
$z(P_v)={3\over 2}$ for $|P_v|=2$; the factor  
$\prod_{v\ {\rm not}\ {\rm e.p.}}
2^{-(h_v-h_{v'})z(P_v)}$ takes into account the presence of the $\RR$ operation
on the vertices.
Once that the bound \pref{2.57aab} is obtained, we have to see if we can sum over the scales and the trees.
Let us define $n(v)=\sum_{i: v_i^*>v}\,1$ as the number of endpoints following
$v$ on $\t$ and $v'$ as
the vertex immediately preceding $v$ on $\t$. Recalling that $|I_v|$
is the number of field labels associated to the endpoints following $v$
on $\t$ (note that $|I_v|\ge 2 n(v)$) and using that
\bea && \sum_{v\ {\rm not}\ {\rm e.p.}}\Big[\big(\sum_{i=1}^{s_v}
|P_{v_i}|\big)-|P_v|\Big]=|I_{v_0}|-|P_{v_0}|\;,\nn\\
&&
\sum_{v\ {\rm not}\ {\rm e.p.}}(s_v-1)=n-1\label{2.58}\\
&& \sum_{v\ {\rm not}\ {\rm e.p.}}
(h_v-h)\Big[\big(\sum_{i=1}^{s_v}
|P_{v_i}|\big)-|P_v|\Big]=\sum_{v\ {\rm not}\ {\rm e.p.}}
(h_v-h_{v'})(|I_v|-|P_v|)\;\nn\\
&&\sum_{v\ {\rm not}\ {\rm e.p.}}(h_v-h)(s_v-1)=
\sum_{v\ {\rm not}\ {\rm e.p.}}(h_v-h_{v'})(n(v)-1)\;,\nn\eea
we find that \pref{2.57aab} can be bounded above by
\bea &&\sum_{n\ge 1}\sum_{\t\in {\cal T}_{h,n}}
\sum_{\PP\in{\cal P}_\t\atop |P_{v_0}|=2l}\sum_{T\in{\bf T}}
C^n  \e_0^n 2^{h({7\over 2}-{5\over 4}|P_{v_0}|+{7\over 4}|I_{v_0}|-{7\over 2}n)} \\
&&
\Big[\prod_{v\ {\rm not}\ {\rm e.p.}} \frac{1}{s_v!}
2^{(h_v-h_{v'})({7\over 2}-{5 |P_v|\over 4}+{5|I_v|\over 4}-{7\over 2}n(v)+z(P_v))}\Big]
\Big[\prod_{v\ {\rm e.p.}, |I_v|=2} 2^{ h_{v'}(1+\d_v)} \Big]\nn
\eea
Using the identities
\bea &&2^{h  n}
\prod_{v\ {\rm not}\ {\rm e.p.}}
2^{(h_v-h_{v'}
) n(v)}=\prod_{v\ {\rm e.p.}}
2^{h_{v'}}\;,\nn\\
&& \g^{h  |I_{v_0}|}
\prod_{v\ {\rm not}\ {\rm e.p.}}
2^{(h_v-h_{v'}) |I_v|}=\prod_{v\ {\rm e.p.}}
2^{h_{v'} |I_v|}\;,\label{2.60}\eea
we finally obtain
\bea&& \frac1{\b |\L|}\int d\xx_1\cdots d\xx_{l}|W^{(h)}_{l}
(\xx_1,\ldots,\xx_{l})|\le \sum_{n\ge 1}\sum_{\t\in {\cal
T}_{h,n}} \sum_{\PP\in{\cal P}_\t\atop |P_{v_0}|=l}\sum_{T\in{\bf
T}} C^n  \e_0^n 2^{h({7\over 2}-{5 l\over 4})} 
\label{2.61}\nn\\
&&\cdot \Big[\prod_{v\ {\rm not}\ {\rm e.p.}}
\frac{1}{s_v!} 2^{-(h_v-h_{v'})(5{|P_v|\over 4}-{7\over
2}+z(P_v))}\Big]\Big[\prod_{v\ {\rm e.p.}} 2^{h_{v'}({5|I_v|/4-7/2})}
\Big]\\
&& \Big[\prod_{v\ {\rm e.p.}, |I_v|=2} 2^{h_{v'}(1+\d_v)} \Big]\nn
\eea
Note that, if $v$ is not an endpoint, $5{|P_v|\over 4}-{7\over
2}+z(P_v)
\ge {1\over 2}$ by the
definition of $\RR$.
%
%
%
Now, note that the number of terms in $\sum_{T\in {\bf T}}$ can be
bounded by $C^n\prod_{v\ {\rm not}\ {\rm e.p.}} s_v!$. Using also
that $5{|P_v|\over 4}-{7\over
2}+z(P_v)\ge 1/2$ and $|P_v|-3\ge|P_v|/4$, we find
that the l.h.s. of (\ref{2.52sz}) can be bounded as
\bea&& \frac1{\b |\L|}\int d\xx_1\cdots d\xx_{l}|W^{(h)}_{l}
(\xx_1,\ldots,\xx_{l})|\le 2^{h({7\over 2}-{5 l\over 4})}
\sum_{n\ge 1}C^n\e_0^n\sum_{\t\in {\cal T}_{h,n}}\cdot\label{2.61b}\\
&&\cdot\big(
\prod_{v\ {\rm not}\ {\rm e.p.}}
2^{-(h_v-h_{v'})/4}\big)
\sum_{\PP\in{\cal P}_\t\atop |P_{v_0}|=2l}\big(\prod_{v\ {\rm not}\ {\rm e.p.}}
2^{-|P_v|/8}\big)
\;.\nn \eea
%
The sum over $\PP$ can be bounded using the following combinatorial
inequality: let $\{p_v, v\in \t\}$,
with $\t\in\TT_{h,n}$, a set of integers such that
$p_v\le \sum_{i=1}^{s_v} p_{v_i}$ for all $v\in\t$ which are not endpoints;
then $\prod_{\rm v\;not\; e.p.} \sum_{p_v} 
2^{-p_v/8}\le C^n$.
%
%
%
Finally
$$\sum_{\t\in {\cal T}_{h,n}}
\prod_{v\ {\rm not}\ {\rm e.p.}}2^{{1\over 2}(h_v-h_{v.})}
\le C^n\;,$$
as it follows by the fact that the number of non trivial vertices in $\t$
is smaller than $n-1$ and that the number of trees in ${\cal T}_{h,n}$ is
bounded by ${\rm const}^n$, and collecting all the previous bounds, we obtain
\pref{2.52sz}.
In order to derive \pref{2.52sz1} we note that, for any tree with no $\n$ end-points
\be
\Big[\prod_{v\ {\rm e.p.}} 
2^{h_{v'}(5|I_v|/4-7/2)}
\Big] \Big[\prod_{v\ {\rm e.p.}, |I_v|=2} 2^{h_{v'}(1+\d_v)} \Big]\le 
\prod_{v\ {\rm e.p.}} 2^{{1\over 2}h_{v'}}
\ee
so that we can replace \pref{2.61} by
\bea&& \frac1{\b |\L|}\int d\xx_1\cdots d\xx_{l}|W^{(h)}_{l}
(\xx_1,\ldots,\xx_{l})|\le \sum_{n\ge 1}\sum_{\t\in {\cal
T}_{h,n}} \sum_{\PP\in{\cal P}_\t\atop |P_{v_0}|=l}\sum_{T\in{\bf
T}} C^n  2^{h({7\over 2}-{5 l\over 4})} 
C^n \e_0^n\cdot\label{2.61a}\nn\\
&&\hskip.2truecm \cdot \Big[\prod_{v\ {\rm not}\ {\rm e.p.}}
\frac{1}{s_v!} 2^{-{1\over 2}(h_v-h_{v'})(5{|P_v|\over 4}-{7\over
2}+z(P_v))}\Big] 
\Big[\prod_{v\ {\rm not}\ {\rm e.p.}}
2^{-{1\over 4}(h_v-h_{v'})}\Big] 
 \prod_{v\ {\rm e.p.}} 2^{{1\over 2}h_{v'}}
\eea
and
\be
\Big[\prod_{v\ {\rm not}\ {\rm e.p.}}
2^{-{1\over 4}(h_v-h_{v'})}\Big] 
 \prod_{v\ {\rm e.p.}} 2^{{1\over 2}h_{v'}}\le 2^{h\over 8}
\ee
This concludes the proof of Lemma 1.
\qed

\subsection{The flow of the effective parameters}

The previous lemma provides convergence of the renormalized expansion
provided that the effective parameters remain close $O(U)$ to their initial value and $U$ is chosen
small enough. 
The flow equation for $\n_h$ can be written as
\be
\n_{h-1}=2^h \n_h+\b^{(h)}_\n
\ee
with, from \pref{2.52sz1}, $\b^{(h)}_\n=O(U 2^{h\over 8})$. By iteration we get
\be
\n_h=2^{-h+1}[\n+\sum_{k=h+1}^1 2^{k-2}\b^{(k)}_\n]\ee
If we choose $\n$ so that 
\be
\n=-\sum_{k=h^*+1}^1 2^{k-2}\b^{(h)}_\n+2^{h^*-1}\n_{h^*}
\ee
then 
\be
\n_h=-2^{-h}\sum_{k=h^*+1}^h 2^{k-2}\b^{(h)}_\n+2^{h^*-h}\n_{h^*}\label{11}
\ee
By a fixed point argument one can prove, see for instance Lemma 4.2 of \cite{BM}, that it is possible 
to find a sequence of $\n_h$ solving \pref{11}. Moreover, $v_{i,h-1}-v_{i,h}=O(U 2^{h/8})$,
${Z_{h-1}\over Z_h}=1+O(U 2^{h/8})$, from 
 \pref{2.52sz1}, so that $v_{i,h-1}-v_{i,0}=O(U)$,
$Z_{h-1}=1+O(U)$.

\section{Renormalization Group integration: the second regime}

\subsection{Tree expansion and convergence}

While the analysis of the scales grater than $h^*$ are insensitive to the sign of $r$, the integration of the 
scales smaller than $h^*$ depends on it. The case $r<0$ 
corresponds to the insulating phase; all the scales $\le h^*$ can be integrated in a single step (setting $\n_{h^*}=0$)
as the 
propagator of $\psi^{(\le h^*)}$ has the same asymptotic behavior of the single scale propagator for $h\ge h^*$, that is
\be
|g^{(\le h^*)}(\xx)|\le {1\over Z_{h^*}}
 {2^{5h^*\over 2}\over 1+[2^{h^*} (|x_0|+|x_+|+|x_-|)+2^{h^*\over 2}|x_3|]^N}
\label{baa}
\ee
Similarly the case $r=0$ correspond to the case $h^*=-\io$ and it can be analyzed as in the previous section.

Let us consider now the case $r>0$, corresponding to the metallic phase:
the Fourier transform of 
the 
propagator vanishes now in correspondence of two Fermi momenta and 
we need a multiscale decomposition. We write
\be
\hat g^{(\le h^*)}(\kk)=\bar g^{(h^*)}(\kk)+\hat g^{(< h^*)}(\kk)
\ee
where $\hat g^{(< h^*)}(\kk)$ is equal to $\hat g^{(\le h^*)}(\kk)$ with
$\chi_{h^*-1}(\kk)$ replaced by 
\be
 \sum_{\o=\pm }\theta(\o k_3) \bar\chi_{h^*-1}(\kk)\label{sasas}
\ee
and $\chi_{h^*-1}(\kk)
=\bar \chi(
b_0 2^{-h^*}|\det A_{h^*}(k)|^{1\over 2})$
with
$b_0$ chosen so that $\chi_{h^*-1}(\kk)$
has support in two disconnected regions around $\pm \pp_F$.
The propagator $\bar g^{(h^*)}(\kk)$, with support in $\chi_{h^*}-\chi_{h^*-1}$, verifies the same bound as \pref{b} with $h=h^*$; in fact
the denominator of $\hat g^{(h^*)}(\kk)$ is $O(r)$; moreover, if $k=k'+\o p_F$,
one has $\cos(k'+\o p_F)-\cos p_F=v_{3,0} k'+{1\over 2} k'^2+O(k'^3)$, 
with $v_{3,0}=O(\sqrt{r})$ for small $r$.
Therefore each derivative with respect to $k'$ produces an extra $r^{-{1\over 2}}$. We can decompose the Grassmann variables
as
\be
\psi^{\pm(\le h^*)}_{\xx}=\psi^{\pm(h^*)}_\xx+\sum_{\o=\pm} e^{\pm i\o \pp_F \xx}\psi^{\pm(<h^*)}_{\o,\xx}
\ee
where $\psi^{\pm(<h^*)}_{\xx}$ has propagator 
\be
g^{(<h^*)}_{\o}(\xx)=\int d\xx e^{i\kk'\xx} \hat g^{(< h^*)}(\kk'+\o\pp_F)
\ee
We can therefore integrate 
$\psi^{(h^*)}$ so that
\bea
&&e^{|\L|\b E_{h^*}}\int P(d\psi^{(h^*)})\int \prod_{\o=\pm} P(d\psi_\o^{(< h^*)})e^{\VV^{(h^*)}(Z_{h^*}\psi^{(\le h^*)})}=\\
&&\int \prod_{\o=\pm} P(d\psi_\o^{(< h^*)})e^{\VV^{(h^*-1)}(Z_{h^*-1}\psi^{(< h^*)})}\nn
\eea
where 
\be \VV^{(h^*-1)}(\psi)=\sum_{n\ge 1}
\int d\xx_1...\int d\xx_n 
W^{(h)}_{n}(\underline\xx)[\prod_{i=1}^n e^{i\e_i\o_i\pp_F\xx_i}\psi^{\e_i(\le h)}_{\o_i,\xx_i}]
\label{coco}\ee
and $W^{(h)}_{n}(\underline\xx)$ is translation invariant. 

We describe the integration of the scales $h< h^*$ inductively. Assume that we have integrate the scale $h^*,-,..,h+1$
showing that \pref{gf} can be written as 
\be
e^{|\L| \b E_h}\int  \prod_{\o=\pm}P(d\psi_\o^{(\le h)}) e^{\tilde \VV^{(h)}(\sqrt{Z_{h}}\psi^{(\le h)})}\label{ef1}
\ee
where $P(d\psi^{(\le h)})$ has propagator given by
%
\bea
&&g^{(\le h)}(\xx)_\o=\\
&&\int d\kk e^{i\kk\xx}{\chi_h(\kk)\over Z_h} 
\begin{pmatrix}&-i k_0+\o v_{3,h}\sin k'_3+E'(\kk)& 
v_{h}(\sin k_+-i\sin k_-)\\
&
v_{h}(\sin k_++i \sin k_-)& 
-i k_0-\o v_{3,h}\sin k'_3- E(kk)
\end{pmatrix}^{-1}\label{cond2}\nn
\eea
%
where $E'(\kk)=\cos p_F (\cos k_3-1)+E(\vec k)$.
and $V^{(h)}(\psi)$ is similar to \pref{coco}
We introduce a {\it localization operator} acting on the effective potential as in \pref{loc}
acting on the kernels 
$\hat W^{(h)}_{n}(\kk_1,..,\kk_{n-1})$ in the following way:
\begin{enumerate}
\item $\LL \hat W^{(h)}_{n}(\kk_1,..,\kk_{n-1})=0$ if $n>2$.
\item If $n=2$
\be
\LL  \hat W^{(h)}_{2}(\kk)= \hat W^{(h)}_{2}(\o\pp_F)+k_0  
\hat W^{(h)}_{2}(\o\pp_F)+\sum_{i=+,-,3} \sin k'_i  
\partial_i \hat W^{(h)}_{2}(\o\pp_F)
\ee
\end{enumerate}
Note that, by symmetry
\bea
&&\hat W^{(h)}_{2}(\o p_F)=\s_3 n_h\quad \partial_+ \hat W^{(h)}_{2}(0)=\s_1 b_{+,h}\nn\\
&&
\partial_- \hat W^{(h)}_{2}(\o p_F)=\s_2 b_{-,h}\quad \partial_3\hat W^{(h)}_{2}(\o p_F)=\s_3 b_{3,h}
\eea
We can include the quadratic part in the free integration; the single scale propagator verifies the following bound
\bea
&&|g^h_\o(\xx)|\le {1\over v_{3,h}}{2^{3h}\over 1+2^h (|
x_0|+|x_+|+|x_-|)+v_{3,h}^{-1} |x|)^N}\nn\\
&&
\int d\xx |g^h(\xx)|\le C 2^h\quad\quad \max |g^h(\xx)|\le {2^h\over v_{3,h}}
\label{x1}\eea
and $v_{3,h^*}=O(\sqrt{r})$. Note also that
\be
g^{(h)}_\o(\xx)=g^{(h)}_{rel,\o}(\xx)+r^{(h)}_{\o}(\xx)\label{dec}
\ee
where
%
\bea
&&g^{(\le h)}(\xx)_\o=\\
&&\int d\kk e^{i\kk\xx}{\chi_h(\kk)\over Z_h} 
\begin{pmatrix}&-i k_0+\o v_{3,h}\sin k'_3& 
v_{h}(\sin k_+-i\sin k_-)\\
&
v_{h}(\sin k_++i \sin k_-)& 
-i k_0-\o v_{3,h}\sin k'_3-
\end{pmatrix}^{-1}\label{cond2}\nn
\eea
%
and $r^{(h)}$ verifies a similar bound with an extra $2^h$.
\begin{lemma}
If $r>0$ there exists a constant $\e_0$ independent of 
of $\b$, $L$ and $r$, such that for $|U|\le \e_0$
and $\max_{k\ge h}[|\n_k|, |Z_k-1|,|{v_{k,i}\over v_{0,i}}-1|\le \e_0$, $i=\pm,3$ then for $h\le h^*$
\be \frac1{\b |\L|}\int d\xx_1\cdots d\xx_{l}|W^{(h)}_{l}
(\xx_1,\ldots,\xx_{l})|\le
2^{h (4-3 l/2)} \,(C \e_0)^{max(1,l/2-1)}\;.\label{2.52sza}\ee
and, if  $\bar W^{(h)}_{l}$ is given  by
\pref{bar} 

\be \frac1{\b |\L|}\int d\xx_1\cdots d\xx_{l}|\bar W^{(h)}_{l}
(\xx_1,\ldots,\xx_{l})|\le
2^{h (4-3 l/2)}2^{{1\over 8} (h-h^*)} \,(C\e_0)^{max(1,l/2-1)}\;.\label{2.52sz1a}\ee
with $C$ a suitable constant.
\end{lemma}
\vskip.3cm
{\it Proof.} 
Again the effective potential can be written as a sum over trees similar to the previous ones
..but with some modifications ( see Fig 2).

\insertplot{300}{150}{ \ins{30pt}{85pt}{$r$} \ins{50pt}{85pt}{$v_0$}
\ins{130pt}{100pt}{$v$} \ins{35pt}{-5pt}{$h$} \ins{52pt}{-5pt}{$h+1$}
\ins{135pt}{-5pt}{$h_{v}$} \ins{215pt}{-5pt}{$-1$} \ins{235pt}{-5pt}{$h^*$}
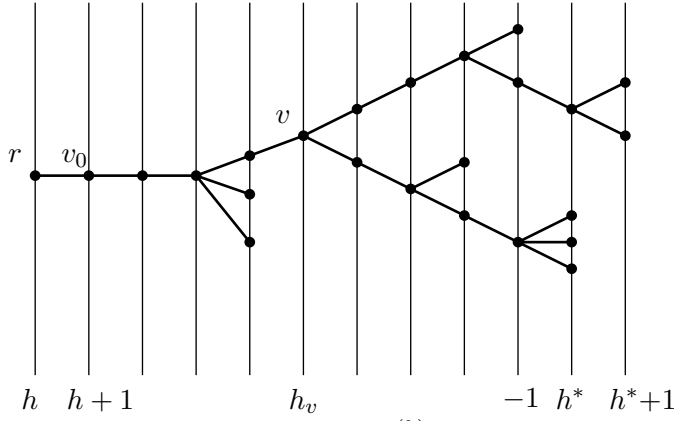
\ins{255pt}{-5pt}{$h^*$+1}} {fig50} {A renormalized tree for
$\VV^{(h)}$ in the second regime\lb{h2aa}}{0}

The scales are
$\le h^*$ and 5) in the previous definition is replaced by:
\vskip.3cm

\0 5') With each endpoint $v$ we associate one of the monomials
with four or more Grassmann fields contributing to $\RR {\cal
V}^{(h^*)}(\psi^{(\le h_v-1)})$ and a set $\xx_v$ of space-time
points (the corresponding integration variables in the $\xx$-space
representation); or a term corresponding to $\LL\VV^{(h_v-1)}$.

In terms of these trees, the effective potential ${\cal V}^{(h)}$, $h\le -1$
is defined as 

i) if $s>1$, then
\be {\cal V}^{(h)}(\t,\psi^{(\le h)})={(-1)^{s+1}\over s!} \EE^T_{h+1}
\big[\bar{\cal V}^{(h+1)}(\t_1,\psi^{(\le h+1)});\ldots; \bar{\cal V}^{(h+1)}
(\t_{s},\psi^{(\le h+1)})\big]\;,\label{2.42}\ee
where $\bar{\cal V}^{(h+1)}(\t_i,\Psi^{(\le h+1)})$ is equal to $\RR{\cal
V}^{(h+1)}(\t_i,\psi^{(\le h+1)})$ if the subtree $\t_i$ contains
more than one end-point, or if it contains one end-point but it is
not a trivial subtree;
it is equal to $\RR{\cal V}^{(h^*)}(\t_i,\Psi^{(\le h+1)})$
or $\g^{h+1}\n_{h+1} F_\n(\psi^{(\le h+1)})$
if $\t_i$ is a trivial subtree;\\
ii) if $s=1$, then ${\bar V}^{(h+1)}(\t,\psi^{(\le h)})$ is equal to
$\big[\RR{\cal V}^{(h+1)}(\t_1,\psi^{(\le h+1)})\big]$
if $\t_1$ is not a trivial
subtree; it is equal to $\big[\RR {\cal
V}^{(h^*)}(\psi^{(\le h+1)})- \RR{\cal V}^{(h^*)}(\psi^{(\le h)})\big]$ or
if $\t_1$ is a trivial subtree.


As before, we
we get 
\be {\cal V}^{(h)}(\t,\PP) = \sum_{T\in {\bf T}} \int d\xx_{v_0}
\widetilde\psi^{(\le h)}(P_{v_0}) W_{\t,\PP,T}^{(h)}(\xx_{v_0})
\= \sum_{T\in {\bf T}}
{\cal V}^{(h)}(\t,\PP,T)\;,\label{2.49}\ee
%
where,
given $\t\in\TT_{h,n}$ and the labels $\PP,T$,
calling $I_R$ the endpoints of $\t$ 
to which is associated $\RR\VV^{(h^*)}$ and 
$I_\n$ the end-points associated to $\LL\VV^{(h_v-1)}$, the explicit representation of $W_{\t,\PP,T}^{(h)}
(\xx_{v_0})$ in (\ref{2.49}) is
\bea &&
W_{\t,\PP, T}(\xx_{v_0}) =\left[\prod_{v\in I_R}
K_{v}^{(h^*)} (\xx_{v})\right]\prod_{v\in I_v}\g^{h_v}\n_{h_v};
\Bigg\{\prod_{v\,\atop\hbox{\ottorm not e.p.}}{1\over s_v!} \int
dP_{T_v}({\bf t}_v)\nn\\
&&\;{\rm det}\, \tilde G^{h_v,T_v}({\bf t}_v)
\Biggl[
\prod_{l\in T_v} \d_{\o^-_l,\o^+_l}\,
\big[(\xx_l-\yy_l)^{\a_l}\partial^{\b_l}g^{(h_v)}_{\o_l}(\xx_l-\yy_l)\big]\,\Biggr]
\Bigg\}\;\eea
where $K_{v}^{(h^*)} (\xx_{v})$ are the kernels of $\RR\VV^{(h^*)}$.
By using the bounds obtained in the previous regime \pref{2.52sz}
\bea 
&&\int\prod_{l\in T^*\setminus\cup_v T_v}d(\xx_l-\yy_l)
\left[\prod_{v\in I_R}K_{v}^{(h^*)} (\xx_{v})\right]\prod_{v\in I_v} 2^{h_v}|\n_{h_v}|
\le\nn\\
&& C^n\e_0^n
\Big[\prod_{v\ {\rm
e.p.}; |I_v|=2} 2^{h_{v'}+\d_v(h_{v'}-2 h^*))}\Big]
\Big[\prod_{v\ {\rm
e.p.}\in I_R, |I_v|\ge  4}2^{h^*({7\over 2}-{5|I_v|\over 4})}\Big]
\eea
where $\d_v=1$ if $v\in I_R$ (again if $v\in I_R$ the factor $2^{2(h_{v'}- h^*)}$ comes from the definition of $\RR$) .
Therefore
\bea && \frac1{\b |\L|}\int d\xx_1\cdots d\xx_{l}|W^{(h)}_{l}
(\xx_1,\ldots,\xx_l)|\le C^n \e_0^n\\
&&\sum_{n\ge 1}\sum_{\t\in {\cal T}_{h,n}}
\sum_{\PP\in{\cal P}_\t\atop |P_{v_0}|=
l}\sum_{T\in{\bf T}}
\Big[\prod_{v\ {\rm not}\ {\rm e.p.}} \frac{1}{s_v!}\nn\\
&&[2^{h_v}]^{\left(\sum_{i=1}^{s_v}{3|P_{v_i}|\over 2}-3{|P_v|\over 2}-4(s_v-1)\right)}
[{1\over v_{3,0}}]^{\left(\sum_{i=1}^{s_v}{|P_{v_i}|\over 2}-{|P_v|\over 2}-(s_v-1)\right)}\Big]\Big]\nn
\\
&&\Big[\prod_{v\ {\rm
e.p.}; |I_v|=2} 2^{h_{v'}+\d_v(h_{v'}-2 h^*))}\Big]
\Big[\prod_{v\ {\rm
e.p.}\in I_R, |I_v|\ge  4}2^{h^*({7\over 2}-{5|I_v|\over 4})} \Big]\nn
\label{2.57a}\nn\eea
%
and 
by using (67)
\bea && \frac1{\b |\L|}\int d\xx_1\cdots d\xx_{l}|W^{(h)}_{l}
(\xx_1,\ldots,\xx_l)|\le C^n \e_0^n\nn\\
&&\sum_{n\ge 1}\sum_{\t\in {\cal T}_{h,n}}
\sum_{\PP\in{\cal P}_\t\atop |P_{v_0}|=
l}\sum_{T\in{\bf T}}
2^{h({4-{3\over 2}|P_{v_0}|+{3\over 2}|I_{v_0}|-4 n)}}\nn\\
&&
\Big[\prod_{v\ {\rm not}\ {\rm e.p.}} \frac{1}{s_v!}
2^{(h_v-h_{v'})(4-{3|P_v|\over 2}+{3|I_v|\over 2}- 4 n(v))}\Big]\\
&&\Big[\prod_{v\ {\rm
e.p.};  |I_v|=2} 2^{h_{v'}(1+\d_v(h_{v'}-2 h^*))}\Big]
\Big[\prod_{v\ {\rm
e.p.}\in I_R, |I_v|\ge 4}2^{h^*({7\over 2}-{5|I_v|\over 4})} \Big]\nn\\
&&\Big[\prod_{v\ {\rm not}\ {\rm e.p.}}
[{1\over v_{3,0}}]^{\left(\sum_{i=1}^{s_v}{|P_{v_i}|\over 2}-{|P_v|\over 2}-(s_v-1)\right)}\Big]]\nn
\eea
and finally using \pref{2.60}
\bea && \frac1{\b |\L|}\int d\xx_1\cdots d\xx_{l}|W^{(h)}_{l}
(\xx_1,\ldots,\xx_l)|\le\\
&&\sum_{n\ge 1}\sum_{\t\in {\cal T}_{h,n}}
\sum_{\PP\in{\cal P}_\t\atop |P_{v_0}|=l}\sum_{T\in{\bf T}}
C^n  \e_0^n 2^{h({4-{3\over 2}|P_{v_0}|)}}\nn\\
&&
\Big[\prod_{v\ {\rm not}\ {\rm e.p.}} \frac{1}{s_v!}
2^{(h_v-h_{v'})(4-{3|P_v|\over 2})}\Big]\Big[\prod_{v\ {\rm
e.p.};  |I_v|=2, v\in I_R} 2^{2(h_{v'}-h^*) }\Big]\nn\\&&
\Big[\prod_{v\ {\rm
e.p.}; v\in I^{R},|I_v|\ge 4} 2^{h_{v'}(-4+{3|I_v|\over 2})} 
 2^{h^*({7\over 2}-{5|I_v|\over 4})} ] \nn\\
&&\Big[\prod_{v\ {\rm not}\ {\rm e.p.}}
[{1\over v_{3,0}}]^{\left(\sum_{i=1}^{s_v}{|P_{v_i}|\over 2}-{|P_v|\over 2}-
(s_v-1)\right)}\Big]]\nn
\eea
By writing
\bea
&&\Big[\prod_{v\ {\rm
e.p.}; v\in I^{R}} 2^{h_{v'}(-4+{3|I_v|\over 2})} 
 2^{h^*({7\over 2}-{5|I_v|\over 4})} ] =\nn\\
&&\Big[\prod_{v\ {\rm
e.p.}; v\in I^{R}} 2^{(h_{v'}-h^*)(-4+{3|I_v|\over 2})} 
 2^{h^*(-{1\over 2}+{|I_v|\over 4})} ] 
\eea
and using that $2^{h^*(-{1\over 2}+{|I_v|\over 4})}\le C (v_{3,0})
^{-1+{|I_v|\over 2}})$
we get
\bea && \frac1{\b |\L|}\int d\xx_1\cdots d\xx_{l}|W^{(h)}_{l}
(\xx_1,\ldots,\xx_l)|\le C^n \e_0^n\\
&&\sum_{n\ge 1}\sum_{\t\in {\cal T}_{h,n}}
\sum_{\PP\in{\cal P}_\t\atop |P_{v_0}|=l}\sum_{T\in{\bf T}}
C^n  \e_0^n \g^{h({4-{3\over 2}|P_{v_0}|)}}\nn\\
&&\Big[\prod_{v\ {\rm not}\ {\rm e.p.}} \frac{1}{s_v!}
2^{(h_v-h_{v'})(4-{3|P_v|\over 2})}\Big]
\Big[\prod_{v\ {\rm
e.p.};  |I_v|=2, v\in I_R} 2^{2(h_{v'}-h^*) }\Big]\nn\\
&&\Big[\prod_{v\ {\rm not}\ {\rm e.p.}}
[{1\over v_{3,0}}]^{\left(\sum_{i=1}^{s_v}{|P_{v_i}|\over 2}-{|P_v|\over 2}-
(s_v-1)\right)}\Big]]\nn\\
&&\Big[\prod_{v\ {\rm
e.p.}, v\in I_R}2^{-(h_{v'}-h^*)(4-{3|I_v|\over 2})} \Big]
\Big[\prod_{v\ {\rm
e.p.}, v\in I_R}
v_{3,0}^{-1+{|I_v|\over 2}}
 \Big]
\label{82}\nn
\eea
Using that
\be
\Big[\prod_{v\ {\rm
e.p.}, v\in I^R} v_{3,0}^{-1+{|I_v|\over 2})}\Big]=\Big[\prod_{v\ {\rm
e.p.}} v_{3,0}^{-1+{|I_v|\over 2})}\Big]
\ee
which follows from the fact that for $v\in I^\n$ one has $|I_v|=2$ so that $
v_{3,0}^{-1+{|I_v|\over 2}}=1$, we can write
\be
\Big[\prod_{v\ {\rm
e.p.}} v_{3,0}^{-1+{|I_v|\over 2})}\Big]\le v_{3,0}^{-n+\sum_{v e.p }|I_v|/2}
\ee
and using $\sum (s_v-1)=n-1$ where $n$ is the number of
end-points we get
\be
\prod_{v\ {\rm
e.p.},} v_{3,0}^{-1}
\prod_{v\ {\rm not}\ {\rm e.p.}} [{1\over v_{3,0}}]^{-
(s_v-1)}\le C v_{3,0}^{-n}
v_{3,0}^{n-1}\le C v_{3,0}^{-1}\ee
Moreover $\sum_{v e.p }|I_v|=l+\sum_v[\sum_{i=1}^{s_v}|P_{v_i}|
-|P_v|]$
\be
\Big[\prod_{v\ {\rm
e.p.}} (v_{3,0})^{|I_v|\over 2}\Big]
\prod_{v\ {\rm not}\ {\rm e.p.}} [{1\over
v_{3,0}}]^{\left(\sum_{i=1}^{s_v}{|P_{v_i}|\over 2}-{|P_v|\over 2}\right)}\Big]\le C
(v_{3,0})^{l/2}\ee
so that in total we get $v_{3,0}^{l/2-1}$ in agreement with
\pref{2.52sza}. Note that the small divisors proportional to $v_{3,0}^{-1}$, which could in principle spoil convergence, are exactly compensated from the factors due to the different scaling of the two regions.
\qed

The flow of the effective coupling can be analyzed as before, noting
that the beta function is $O(U 2^{h-h^*})$ by the above estimate and we get
\bea
&&Z_{h}\to_{h\to-\io} Z=1+O(U^2)\label{aa}\\
&&
v_{3,h}\to_{h\to-\io} v_{3}=t_\perp\sin(p_F)+a_{3} U+O(U^2)
\nn\\
&&v_{\pm, h-1}\to_{h\to-\io} v_{\pm}=t+a_\pm U+O(U^2)\nn\\
\eea
where
\be
a_{3}\s_3=\int d\kk \hat v(\kk) \partial_3 \hat g (\kk)\quad a_{+}\s_1=\int d\kk \hat v(\kk) \partial_+ \hat g(\kk)
\ee
Moreover 
\be
\n=U v(0) \hat S_0(0,0^-)+\int d\kk v(\kk) \hat g(\kk)
\ee
and this concludes the proof of Lemma 2.\qed

From Lemma 1 and Lemma 2 the proof of the main theorem follows easily.



\end{document}